\documentclass[journal,twoside]{IEEEtran}
\usepackage{amsmath,amsfonts}
\usepackage{algorithmic}
\usepackage{algorithm}
\usepackage{array}
\usepackage[caption=false,font=normalsize,labelfont=sf,textfont=sf]{subfig}
\usepackage{textcomp}
\usepackage{stfloats}
\usepackage{url}
\usepackage{verbatim}
\usepackage{graphicx}
\usepackage{cite}
\usepackage{upgreek}
\usepackage{multirow}
\usepackage[none]{hyphenat}
\usepackage{float}
\usepackage{subfig}
\usepackage{adjustbox}
\usepackage{t1enc}
\usepackage{times}
\usepackage{booktabs}
\usepackage{pifont}
\usepackage{booktabs}
\usepackage{tabularx}
\usepackage{array}
\usepackage{arydshln}
\usepackage{tabularray}
\usepackage{makecell}
\usepackage{xcolor}

\newcommand\red[1]{\textcolor{black}{#1}}

\begin{document}

\title{HDR-DTCP: High-Dynamic-Range Energy Harvester Maintaining >50\% PCE with Differential Tissue-Coupled Powering for Reconfigurable Biphasic Stimulation}
\title{HDR-DTCP: High-Dynamic-Range Differential Tissue-Coupled Wireless Powering for Tunable Biphasic Stimulation in Injectable Implants}

\author{Asif~Iftekhar~Omi,~Anyu~Jiang,~Han~Wu,~Sultan~Mahmud,~Mali~Halac,~\IEEEmembership{Graduate Student Member,~IEEE,}\\~Shreyas~Sen,~Shriya~Srinivasan,~Adam~Khalifa,~\IEEEmembership{Senior~Member,~IEEE},~and~Baibhab~Chatterjee,~\IEEEmembership{Member,~IEEE}
\thanks{Asif Iftekhar Omi, Anyu Jiang, Han Wu, Sultan Mahmud, Adam Khalifa, and Baibhab Chatterjee are with the Department of Electrical and Computer Engineering, University of Florida, Gainesville, FL 32611 USA (e-mail: \{as.omi, chatterjee.b\}@ufl.edu).}
\thanks{Shreyas Sen is with the School of Electrical and Computer Engineering, Purdue University, West Lafayette, IN 47907 USA.}
\thanks{Mali Halac and Shriya Srinivasan are with the Department of Bioengineering, Harvard University, Cambridge, MA 02138 USA.}

 }





\maketitle

\begin{abstract}
Next-generation implantable neurostimulators require efficient wireless power delivery in smaller, less invasive form factors, yet variations in tissue coupling, implantation conditions, and stimulation loads can alter received power (or requirements) and degrade energy-harvesting performance.
\red{This work presents high-dynamic-range (HDR) differential tissue-coupled
powering (DTCP), termed HDR-DTCP,}
a body-coupled architecture in which noninvasive differential electrodes establish an AC electric field in tissue and a thread-like implant extracts power from the longitudinal potential gradient for on-chip harvesting and stimulation. Distributed coupling along the implant length, rather than receiver cross-section, enables minimally invasive injectable geometries, with harvested power scaling with implant length.
\red{A 65-nm complementary metal--oxide--semiconductor (CMOS) prototype}
integrates a reconfigurable four-stage energy harvester with automatic stage
selection and a tunable constant-current biphasic stimulator, achieving
63.5\% peak power-conversion efficiency (PCE) at $-16.3$~dBm and
$>50\%$ PCE over a 9-dB input-power range.
Preliminary \textit{in-vivo} validation of DTCP was done with a second \red{application-specific integrated
circuit (ASIC)} on an injectable flexible substrate, capable of being delivered through a 14-gauge needle adjacent to the sciatic nerve of Sprague-Dawley rats (n=4). External copper-pad DTCP powering evoked repeatable hind-limb twitches and \red{electromyography (EMG)} response; ultrasound confirmed sub-millimeter nerve proximity, video-based paw-tip tracking captured the response, and 8-day post-implant dissection verified placement along the sciatic-nerve trajectory.
Overall, this work provides an integrated wireless powering, energy-harvesting, and stimulation platform toward fully battery-free injectable neuromodulation implants.

\end{abstract}

\section{Introduction}
\begin{figure}[t!]
\centerline{\includegraphics[width=1\columnwidth]{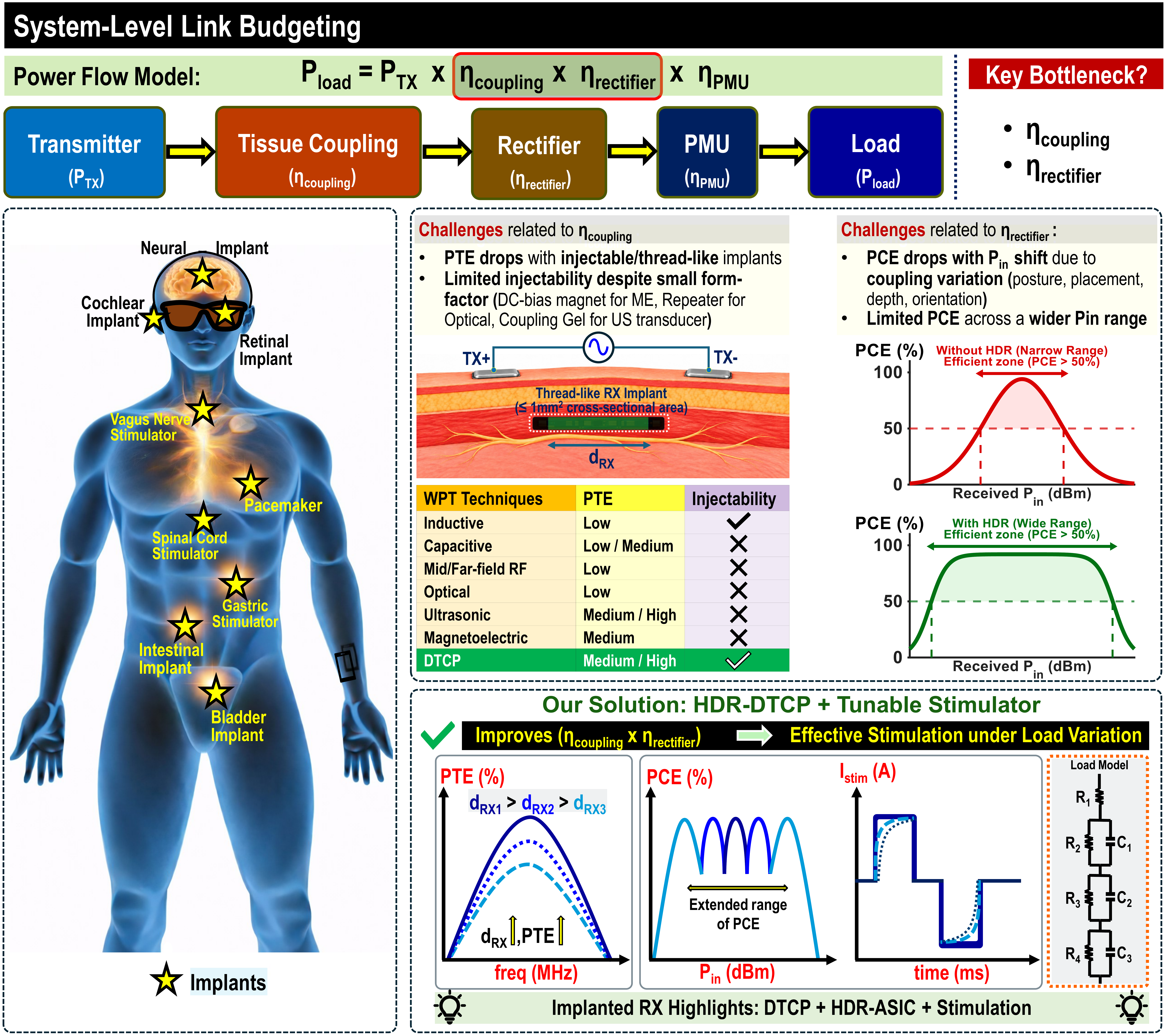}}
\caption{\red{System-level motivation for HDR-DTCP in injectable thread-like biomedical implants. Wireless power delivery depends on tissue-coupling efficiency, $\eta_{\mathrm{coupling}}$, rectifier efficiency, $\eta_{\mathrm{rectifier}}$, and variations in coupling, tissue, and electrode--tissue conditions. HDR-DTCP combines length-scalable coupling, wide-range harvesting, and tunable stimulation to sustain operation across these variations.}}
\label{fig1}
\end{figure}

Emerging implantable bioelectronics are moving toward highly miniaturized form factors, enabling more localized and less invasive interfaces with anatomically constrained biological targets \cite{nair2023miniature,chatterjee2023bioelectronic}. For neuromodulation in particular, reducing implant cross section can facilitate placement near targeted neural structures while limiting tissue disruption. At the same time, this scaling places increasingly stringent demands on wireless power delivery and on the usable on-chip power available to the implant, even with improved sensing, communication and stimulation interfaces \cite{lee2021neural,yu2020implantable,wang2024omnidirectional,lu2025dualoutput,Chatterjee_ISSCC22, Chatterjee_RFIC20,lee2024stimote,zou2025parallel,108483011,chatterjee2021biphasic, Chatterjee_VLSI, Chatterjee_ESSERC22}. Percutaneous connections introduce infection and long-term reliability concerns, whereas batteries occupy substantial implant volume, impose a finite operating lifetime, and may require surgical replacement~\cite{liao2025implantable,singer2021wireless}. Wireless power transfer (WPT) therefore remains a key enabler for battery-free implantable systems~\cite{agarwal2017wireless,zhang2018wireless,monti2021wireless}. As receiver dimensions shrink, however, the power-delivery problem extends beyond transferring energy through tissue: sufficient power must ultimately reach the load within a constrained receiver geometry and under variable implantation and operating conditions. As illustrated in Fig.~\ref{fig1}, 
the load-power budget is jointly constrained by tissue-coupling efficiency and AC-to-DC conversion efficiency, forming the two primary link- and receiver-level bottlenecks.

\subsection{Need for Efficient Coupling and HDR Operation}
The two bottlenecks identified above become increasingly coupled as the implant receiver is miniaturized. At the wireless-link level, $\eta_{\mathrm{coupling}}$ depends strongly on receiver geometry, implantation depth, and alignment. As summarized in Fig.~\ref{fig1}, established WPT modalities trade \red{power transfer efficiency (PTE)} against form factor, misalignment tolerance, depth, and power delivered to the load. Inductive links require sufficiently large and well-aligned coils \cite{omi2024new,khalifa2019microbead}. Capacitive links are strongly influenced by electrode/plate geometry, tissue loading, and the return path formation \cite{koruprolu2018capacitive,jiang2021efficient},\red{\cite{Hossain2022Capacitive}}. Implanted RF antennas suffer reduced efficiency when their dimensions are electrically small and they operate within lossy tissue \cite{das2017multiband}. Ultrasonic links can support compact implants at greater depths, but require acoustic transducers, coupling, and suitable alignment \cite{ghanbari2019sub,Piech2020UltrasonicStimulator},\red{\cite{Kashani2023Ultrasonic}}. Optical links are constrained by tissue absorption and scattering \cite{moon2021bridging},\red{\cite{Chen2017Solar}}, whereas magnetoelectric links commonly rely on specialized material stacks or external bias magnets \cite{alrashdan2021wearable,hosur2023comparative,Chen2022MEImplant},\red{\cite{Mukherjee2024Magnetoelectric}}.

These constraints become particularly restrictive for thread-like receivers, where increasing lateral dimensions can compromise injectability. 
Body-coupled powering (BCP) instead uses tissue as part of the distributed coupling medium rather than treating it only as a lossy channel \cite{li2021body,yoo2023body,li2020body,tudela2021volume,garcia2022wireless,minguillon2022powering,jay2026cicc,107982611,chen2018tissue,107828971,zhao2022ionic,raja2025closed,Shen2026PWMPassiveBCC,Han2026TissueCouplingBCP,Shen2024GalvanicallySwitchingBCC,Han2025MultipleImplants,omi2026IMS,Callejon2012DistributedIBC,Han2024FBLC,Han_adv_sci,Malik2020InjectableSensors,TudelaPi2020VolumeConduction,10782968,han2025ner,9575020,9779520}. Differential tissue-coupled powering (DTCP) represents a differential realization of this principle in which external differential electrodes establish the tissue field while an implanted differential electrode pair extracts power from the resulting spatial potential gradient.
\red{This provides a distinct geometric advantage: increasing the implanted
electrode separation along the receiver length can enhance PTE without
increasing cross-sectional area, thereby preserving a narrow geometry that
can facilitate injectability and more minimally invasive implantation.}

Efficient coupling alone, however, does not ensure sufficient usable DC power. As also highlighted in Fig.~\ref{fig1}, the received AC signal must undergo AC-to-DC conversion, making $\eta_{\mathrm{rectifier}}$ the second critical bottleneck. Rectifier peak \red{power conversion efficiency (PCE)} is inherently input-power dependent, while $P_{\mathrm{in}}$ varies in practice with posture, electrode placement, implantation depth, receiver orientation, tissue impedance, the electrode--tissue interface, and stimulation load, as illustrated in Fig.~\ref{fig1}. Consequently, peak PCE alone is insufficient; high efficiency must be sustained over a broad $P_{\mathrm{in}}$ range to preserve the on-chip power budget under changing operating conditions. This motivates the exploration of DTCP as a differential BCP approach for length-scalable efficient tissue coupling, together with an HDR energy harvester that maintains high $\eta_{\mathrm{rectifier}}$ across the resulting $P_{\mathrm{in}}$ variation---the two complementary directions pursued here.

\subsection{Related Prior Work and Research Gaps}
\newcommand{\TCell}[2]{%
\parbox[t]{#1}{\centering\strut #2\strut}%
}

\begin{table*}[t!]
\centering
\caption{Qualitative Comparison of Representative Prior Works Relevant to HDR-DTCP}
\label{tab:qualitative_comparison}

\renewcommand{\arraystretch}{1.30}
\setlength{\tabcolsep}{5pt}

\resizebox{\textwidth}{!}{%
\begin{tabular}{@{}ccccccc@{}}
\hline

\TCell{2.7cm}{\textbf{Work}} &
\TCell{2.5cm}{\textbf{Primary Bottleneck}} &
\TCell{2.1cm}{\textbf{Technique}} &
\TCell{4.0cm}{\textbf{Performance-Enhancement}\\\textbf{Mechanism}} &
\TCell{3.2cm}{\textbf{On-Chip Adaptation}} &
\TCell{3.0cm}{\textbf{Implant Function}} &
\TCell{2.0cm}{\textbf{Validation}} \\

\hline

\TCell{2.7cm}{\cite{chen2018tissue}\\[-1pt]{\scriptsize TPEL'18}} &
\TCell{2.5cm}{$\upeta_{\mathrm{coupling}}$} &
\TCell{2.1cm}{TCH-WPT} &
\TCell{4.0cm}{Tissue-mediated power transfer} &
\TCell{3.2cm}{---} &
\TCell{3.0cm}{Power delivery} &
\TCell{2.0cm}{Ex vivo} \\

\hline

\TCell{2.7cm}{\cite{minguillon2022powering}\\[-1pt]{\scriptsize TBME'23}} &
\TCell{2.5cm}{$\upeta_{\mathrm{coupling}}$} &
\TCell{2.1cm}{HF-VC} &
\TCell{4.0cm}{Volume-conduction powering} &
\TCell{3.2cm}{---} &
\TCell{3.0cm}{Power delivery} &
\TCell{2.0cm}{In vivo} \\

\hline

\TCell{2.7cm}{\cite{Han2024FBLC}\\[-1pt]{\scriptsize TPEL'24}} &
\TCell{2.5cm}{$\upeta_{\mathrm{coupling}}$} &
\TCell{2.1cm}{CC-IBPT} &
\TCell{4.0cm}{Full-band loss compensation} &
\TCell{3.2cm}{---} &
\TCell{3.0cm}{Power delivery} &
\TCell{2.0cm}{Ex vivo} \\

\hline

\TCell{2.7cm}{\cite{Shen2026PWMPassiveBCC}\\[-1pt]{\scriptsize TBCAS'26}} &
\TCell{2.5cm}{$\upeta_{\mathrm{coupling}}$} &
\TCell{2.1cm}{TCH-WPT} &
\TCell{4.0cm}{Tissue-channel powering} &
\TCell{3.2cm}{---} &
\TCell{3.0cm}{Neural recording} &
\TCell{2.0cm}{In vivo} \\

\hline

\TCell{2.7cm}{\cite{choo2022reconfigurable}\\[-1pt]{\scriptsize TVLSI'22}} &
\TCell{2.5cm}{$\upeta_{\mathrm{rectifier}}$} &
\TCell{2.1cm}{Reconfig.\\Dickson} &
\TCell{4.0cm}{PCE-range extension} &
\TCell{3.2cm}{\red{6/12-stage reconfiguration}} &
\TCell{3.0cm}{Energy harvesting} &
\TCell{2.0cm}{Bench-top} \\

\hline

\TCell{2.7cm}{\red{\cite{Engur2021TripleMode}\\[-1pt]{\scriptsize TBCAS'21}}} &
\TCell{2.5cm}{\red{$\upeta_{\mathrm{rectifier}}$}} &
\TCell{2.1cm}{\red{Triple-mode\\rectifier}} &
\TCell{4.0cm}{\red{Coupling-range extension}} &
\TCell{3.2cm}{\red{Current/half-wave/\\voltage-mode selection}} &
\TCell{3.0cm}{\red{Energy harvesting}} &
\TCell{2.0cm}{\red{Bench-top}} \\

\hline

\TCell{2.7cm}{\cite{choo2023high2}\\[-1pt]{\scriptsize TCAS-II'23}} &
\TCell{2.5cm}{$\upeta_{\mathrm{rectifier}}$} &
\TCell{2.1cm}{CCDD/\\Dickson TA} &
\TCell{4.0cm}{PCE-range extension} &
\TCell{3.2cm}{Mixed Topology} &
\TCell{3.0cm}{Energy harvesting} &
\TCell{2.0cm}{Bench-top} \\

\hline

\TCell{2.7cm}{\red{\cite{Ahn2025AdaptiveDelay}\\[-1pt]{\scriptsize TBCAS'25}}} &
\TCell{2.5cm}{\red{$\upeta_{\mathrm{rectifier}}$}} &
\TCell{2.1cm}{\red{Active\\rectifier}} &
\TCell{4.0cm}{\red{PCE/VCR enhancement}} &
\TCell{3.2cm}{\red{On/off timing tracking}} &
\TCell{3.0cm}{\red{Energy harvesting}} &
\TCell{2.0cm}{\red{Bench-top}} \\

\hline

\TCell{2.7cm}{\textbf{This Work}} &
\TCell{2.5cm}{
\textbf{$\upeta_{\mathrm{coupling}}$}\\[-1pt]
\textbf{+$\upeta_{\mathrm{rectifier}}$}
} &
\TCell{2.1cm}{\textbf{HDR-DTCP}} &
\TCell{4.0cm}{
\textbf{RX-length scaling +}\\
\textbf{wide-PCE harvesting}
} &
\TCell{3.2cm}{
\textbf{Self-bias + \red{PCE-guided}}\\
\textbf{\red{stage selection}}
} &
\TCell{3.0cm}{
\textbf{Tunable biphasic}\\
\textbf{stimulation}
} &
\TCell{2.0cm}{\textbf{Bench-top +\\In vivo\textsuperscript{*}}} \\

\hline
\end{tabular}%
}

\vspace{2pt}

\begin{minipage}{0.98\textwidth}
\footnotesize
\textit{Abbreviations:}
TCH-WPT: tissue-channel wireless power transfer;
HF-VC: high-frequency volume conduction;
CC-IBPT: capacitive-coupled intra-body power transfer;
SBG: self-bias generation;
CCDD: cross-coupled differential drive;
TA: topology amalgamation;
\red{VCR: voltage conversion ratio.}\\
*Bench-top characterization refers to the HDR-DTCP ASIC; in vivo
validation and injectability refer to a separate DTCP ASIC.
\end{minipage}

\end{table*}

Prior BCP studies have established the feasibility of using biological tissue as the power-transfer medium, but have primarily emphasized link feasibility, channel configuration, or application-specific operation. \red{Because several contributions share similar coupling principles and system objectives, representative examples were selected to span distinct BCP directions and provide sufficiently complete quantitative or experimental benchmarks.} An early tissue-channel WPT implementation demonstrated 0.39\% PTE for a 1-mm implant at 5-cm depth in porcine tissue \cite{chen2018tissue}, while subsequent volume-conduction analysis predicted milliwatt-level power extraction for sub-mm implants under safety-constrained excitation \cite{TudelaPi2020VolumeConduction}. Human experiments later demonstrated tissue-mediated delivery of $5.9\pm0.7$~mW in the arm and $2.4\pm0.3$~mW in the lower leg \cite{minguillon2022powering}. More recent work improved capacitive intrabody transfer through loss compensation, achieving 2.06\% PTE with a $1~\mathrm{mm}^{2}$ implanted electrode at 4.6-cm depth \cite{Han2024FBLC}, whereas other studies demonstrated battery-free neural operation or distributed multi-receiver powering through body-channel links \cite{Shen2026PWMPassiveBCC,Han2026TissueCouplingBCP}. These works collectively validate BCP as a viable powering mechanism, yet they do not directly exploit the longitudinal dimension of a thread-like receiver as a systematic means of enhancing PTE while preserving cross-sectional area, nor do they jointly address the receiver-side conversion efficiency
required to utilize the received power.

At the receiver, substantial effort has focused on extending the
input-power range of \red{complementary metal--oxide--semiconductor (CMOS) }rectifiers
\cite{chong2019wide,choo2022reconfigurable,teo2024recent,choo2023high,Li2015Reconfigurable,Erfani2020DualOutput,Engur2021TripleMode,Namgoong2021Calibrated,Ahn2025AdaptiveDelay,choo2023high2,wang2025rf}.
These approaches broadly include reconfigurable structures, multi-path selection,
and self-bias generation \cite{teo2024recent}. Chong \textit{et al.} employed
self-body biasing and leakage suppression in a cross-coupled differential drive (CCDD) rectifier to sustain
efficiency over varying input power \cite{chong2019wide}. Choo \textit{et al.}
introduced a reconfigurable Dickson rectifier switching between 6- and 12-stage
configurations to broaden the input-power range \cite{choo2022reconfigurable}.
Subsequent dual-topology and topology-amalgamation approaches combined the high
forward conduction of CCDD rectifiers with the lower reverse leakage of Dickson
structures to extend the operating range \cite{choo2023high,choo2023high2}.
\red{Other adaptive approaches modify circuit parameters to maintain efficiency
under received-power, coupling, or load variations. Li \textit{et al.} combined
a reconfigurable resonant regulating rectifier with primary equalization to
extend the coupling and loading range \cite{Li2015Reconfigurable}. Erfani
\textit{et al.} employed pulse-width modulation (PWM) and dual-mode
pulse-frequency modulation (PFM) in a single-stage dual-output regulating
rectifier \cite{Erfani2020DualOutput}, whereas Engur \textit{et al.} selected
among current-, half-wave-, and voltage-mode rectification according to the
received-signal voltage \cite{Engur2021TripleMode}. Namgoong \textit{et al.}
calibrated the power-switch transition timing and effective switch size
\cite{Namgoong2021Calibrated}, while Ahn \textit{et al.} used digital-feedback
adaptive delay control to track optimum switch on/off timing under input-voltage
and load variations \cite{Ahn2025AdaptiveDelay}.}
More recently, a wide-range power-harvesting chipset for a wireless
neural-recording headstage combined an RF harvester with a DC--DC booster to
accommodate an input-power range \cite{wang2025rf}. These works demonstrate
circuit-level approaches for maintaining rectifier operation under
varying received power. However, reported power dynamic range (PDR) depends on
the PCE threshold, circuit architecture, load, and application requirements,
making comparison based only on peak PCE or PDR insufficient.

Collectively, previous work has largely addressed tissue-mediated coupling and
wide-range rectification as separate design problems. As summarized in
Table~\ref{tab:qualitative_comparison}, BCP studies have focused primarily on
power-transfer feasibility, channel configuration, and coupling enhancement,
whereas adaptive rectifier studies have mainly addressed receiver-side
conversion over varying input-power conditions. A system that jointly leverages the longitudinal dimension of a thread-like receiver to enhance DTCP power pickup, adapts the energy-harvesting architecture to the resulting received-power variation, and directly supports tunable stimulation within an injectable platform has not yet been demonstrated. These are the coupled research gaps addressed in this work.

\subsection{Proposed HDR-DTCP and  Contributions}
\begin{figure}[t!]
\centerline{\includegraphics[width=0.99\columnwidth]{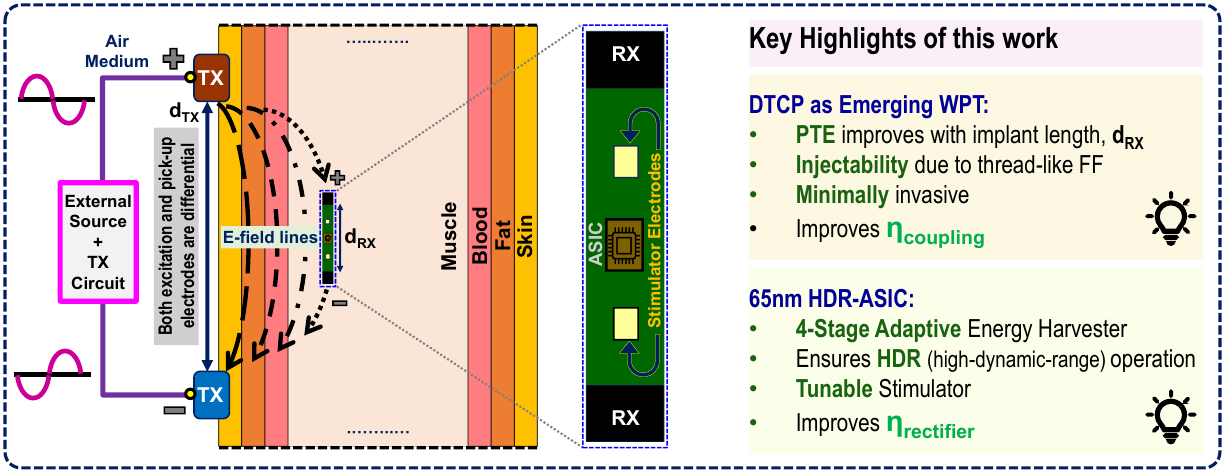}}
\caption{\red{Proposed HDR-DTCP platform for miniaturized injectable implants.
A differential TX establishes the tissue electric field, while the
thread-like RX extracts power from the longitudinal potential gradient.
The 65-nm CMOS ASIC integrates adaptive four-stage energy harvesting and
tunable biphasic stimulation for length-scalable coupling and wide-range
rectification.}}
\label{fig4}
\vspace{-2mm}
\end{figure}

This paper presents high-dynamic-range differential tissue-coupled powering (HDR-DTCP) as an integrated architecture for wireless power delivery, energy harvesting and stimulation in thread-like battery-free implants, as introduced at the system level in Fig.~\ref{fig1} and detailed in Fig.~\ref{fig4}.
A noninvasive differential \red{transmitter (TX)} electrode pair establishes an AC electric field in tissue, while the implanted differential \red{receiver (RX)} extracts energy from the longitudinal potential gradient along the receiver, as illustrated in Fig.~\ref{fig4}. By exploiting the axial dimension of the implant for differential power pickup, DTCP can enhance tissue coupling without increasing receiver cross section. The resulting AC signal is processed by an \red{adaptive four-stage self-bias-generation cross-coupled differential-drive (SBG-CCDD) energy harvester} and used to support on-chip constant-current biphasic stimulation.

 The key contributions are summarized as follows.

\begin{itemize}
\item \textbf{\red{DTCP Framework for Thread-Like Implants}:}
A field-based analysis framework is developed to relate tissue properties,
implantation depth, and TX/RX geometry to differential power transfer. By exploiting the longitudinal dimension of the implant,
DTCP can enhance differential pickup without increasing receiver cross
section, while TX/RX geometry co-design identifies the practical
coupling conditions for narrow, elongated injectable receivers.

\item \textbf{High-Dynamic-Range Wireless Energy Harvesting:}
\red{To accommodate the received-power variation arising from tissue coupling,
implantation depth, and TX/RX geometry,} a four-stage
reconfigurable energy harvester is implemented in 65-nm CMOS with automatic
stage selection to adapt the effective conversion topology to the received
input power. \red{Stage transitions are determined using PCE-derived
reference thresholds, complementing the input-adaptive self-biasing within
each rectifier stage.} The architecture extends efficient AC-to-DC
conversion across the input-power variation encountered in the
tissue-coupled link.

\item \textbf{Tunable Constant-Current Biphasic Stimulation:}
The harvested and regulated supply powers an on-chip tunable biphasic stimulator designed for constant-current operation across the electrode-loading conditions encountered at the stimulation interface.

\item \textbf{Complementary Injectable DTCP Demonstration:}
A separate 65-nm DTCP application-specific integrated circuit (ASIC) assembled on a flexible substrate is delivered through a needle adjacent to the sciatic nerve. The device is capable of being operated wirelessly through DTCP to evoke repeatable neuromuscular responses. Ultrasound imaging, video-based motion tracking, and post-implant dissection provide complementary validation of implant placement and end-to-end system operation.

\end{itemize}

Building on the preliminary HDR-DTCP implementation
\cite{omi2026cicc}, this paper substantially expands the scope through:
\textit{(1) \red{a comprehensive investigation of the underlying DTCP coupling
physics, including tissue-dispersion effects, the spatial-potential-gradient formulation, receiver-length scaling,
and full-wave E-field distributions for different TX/RX geometries
and implantation depths;} (2) \red{an expanded specific absorption rate (SAR)-based safety
and thermal analysis for representative DTCP geometries under
tissue-exposure constraints;} (3) detailed
transistor-level analysis of the SBG-CCDD rectifier, including its
input-adaptive self-bias mechanism and forward- and reverse-conduction
behavior; (4) \red{detailed characterization of the automatic rectifier stage-selection mechanism, including the comparison logic, PCE-derived transition criteria}, stage-dependent PCE characteristics, and adaptive-control flow; and
(5) \red{electrochemical impedance spectroscopy (EIS)}-based characterization
and equivalent-circuit modeling of the stimulation-electrode impedance,
followed by evaluation of stimulator performance under scaled
electrode-load model.}

\subsection{Organization of the Paper}
\label{subsec:paper_organization}
The remainder of this paper is organized as follows. Section II describes the DTCP operating principle, including implant-length-dependent PTE enhancement and SAR-constrained wireless powering. Then, it explains the 65-nm CMOS HDR energy harvester, automatic stage-configuring mechanism, on-chip power conditioning, and digitally controlled constant-current biphasic stimulator. Section III reports the experimental results, including bench-top
HDR energy-harvesting and tunable-stimulation measurements, followed by
in vivo injectable DTCP demonstrations using a separate ASIC and comparing the proposed HDR-DTCP receiver with the state of the art. Section IV concludes the paper.

\section{Design Details of HDR-DTCP Receiver}
\subsection{System Architecture}
\begin{figure}[t!]
\centerline{\includegraphics[width=1\columnwidth]{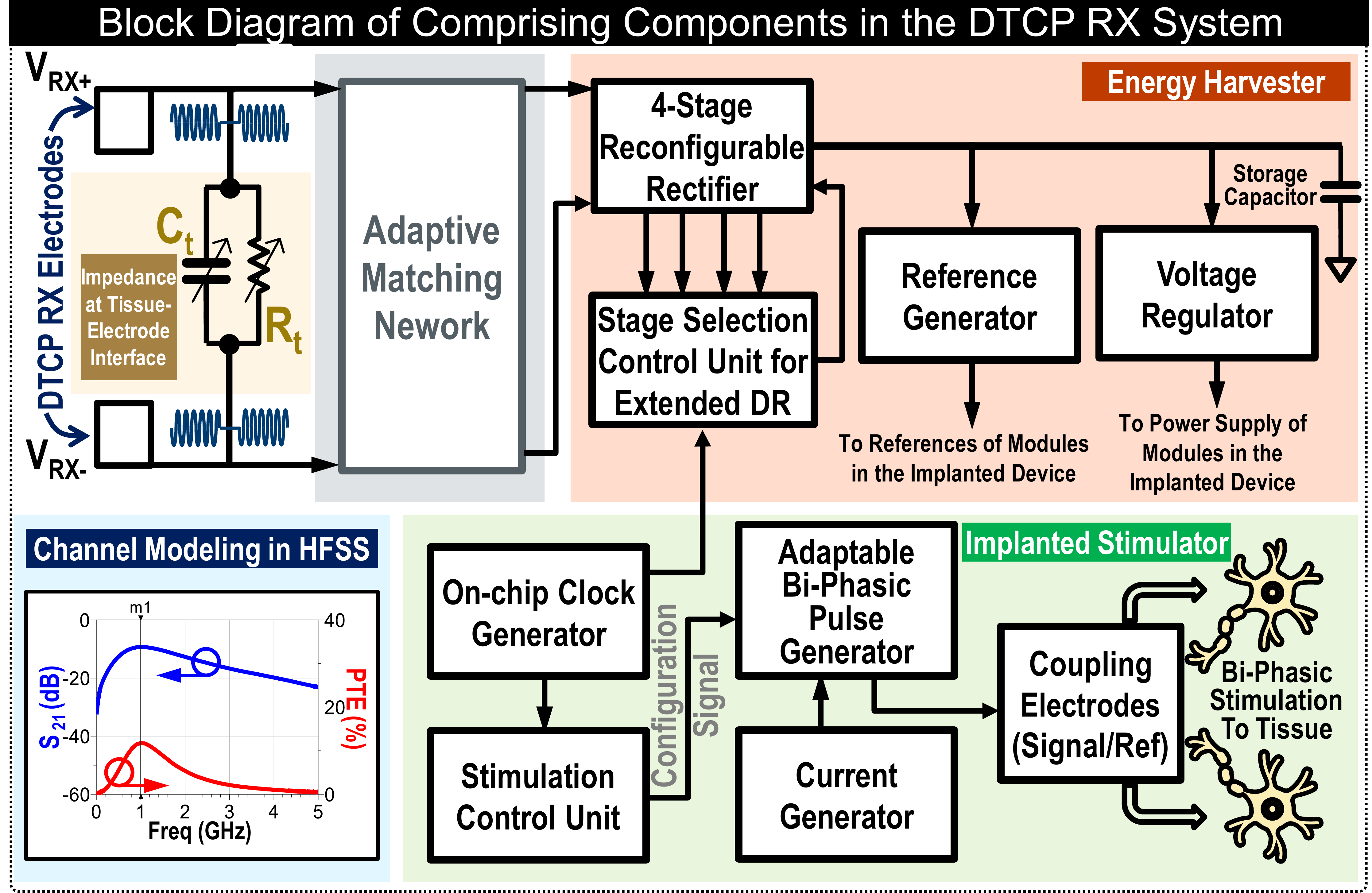}}
\caption{Block-level architecture of the HDR-DTCP receiver. The differential DTCP input is coupled through the tissue--electrode interface and matching network to the four-stage reconfigurable rectifier and stage-selection controller, followed by energy storage, reference generation, and voltage regulation. The regulated supply powers the on-chip clock, stimulation-control, current-generation, and biphasic-pulse circuitry for programmable stimulation.}
\label{ckt0}
\end{figure} 

The HDR-DTCP receiver architecture builds on the preliminary implementation in \cite{omi2026cicc} and integrates the tissue-coupled input interface, adaptive matching, reconfigurable energy harvesting, on-chip regulation, and programmable biphasic stimulation, as shown in Fig.~\ref{ckt0}. The differential DTCP electrodes provide $V_{\mathrm{RX}+}$ and $V_{\mathrm{RX}-}$ through the tissue--electrode interface to the adaptive matching network \cite{wu2024adaptive}, followed by a four-stage reconfigurable rectifier. Intermediate rectifier outputs are monitored by the stage-selection control unit, while the selected $V_{\mathrm{rect}}$ is buffered by the storage capacitor and supplied to the reference-generation and voltage-regulation blocks. The regulated output provides the internal supply for the implant circuitry. The stimulation path comprises an on-chip clock generator, stimulation-control unit, current generator, and adaptable biphasic pulse generator driving the implanted stimulation electrodes. Configuration information received through the DTCP link programs stimulation parameters before the regulated supply is used to generate the commanded biphasic waveform.
A detailed frequency-dependent study of the DTCP link was presented
in \cite{omi2026IMS}, identifying the sub-GHz regime as favorable for
tissue-coupled power transfer. Guided by this prior study, 900~MHz is
selected as the operating frequency for the DTCP link analysis and the
HDR-ASIC design and characterization presented in this work.

\subsection{DTCP Concept and Operating Principle}

\red{As introduced in Fig.~\ref{fig4}, DTCP uses differential external TX and
implanted RX electrode pairs, with $d_{\mathrm{TX}}$ and
$d_{\mathrm{RX}}$ denoting their respective separations and $d$ the
implantation depth. The TX establishes an electric field in the
heterogeneous tissue, while the RX extracts the resulting longitudinal
potential difference.}
Thus,
\begin{equation}
\mathbf{J}
=
\left(
\sigma_{\mathrm{eff}}
+
j\omega\varepsilon_{\mathrm{eff}}
\right)\mathbf{E},
\label{eq:dtcp_current}
\end{equation}

where $\mathbf{J}$ is the tissue current density,
$\sigma_{\mathrm{eff}}$ and $\varepsilon_{\mathrm{eff}}$ are the
effective tissue conductivity and permittivity, respectively, $\omega$
is the angular frequency, and $\mathbf{E}$ is the local electric field.
The conductive term $\sigma_{\mathrm{eff}}\mathbf{E}$ and
displacement-current term
$j\omega\varepsilon_{\mathrm{eff}}\mathbf{E}$ jointly determine the
tissue-field distribution.

\begin{figure}[t!]
\centerline{\includegraphics[width=1\columnwidth]
{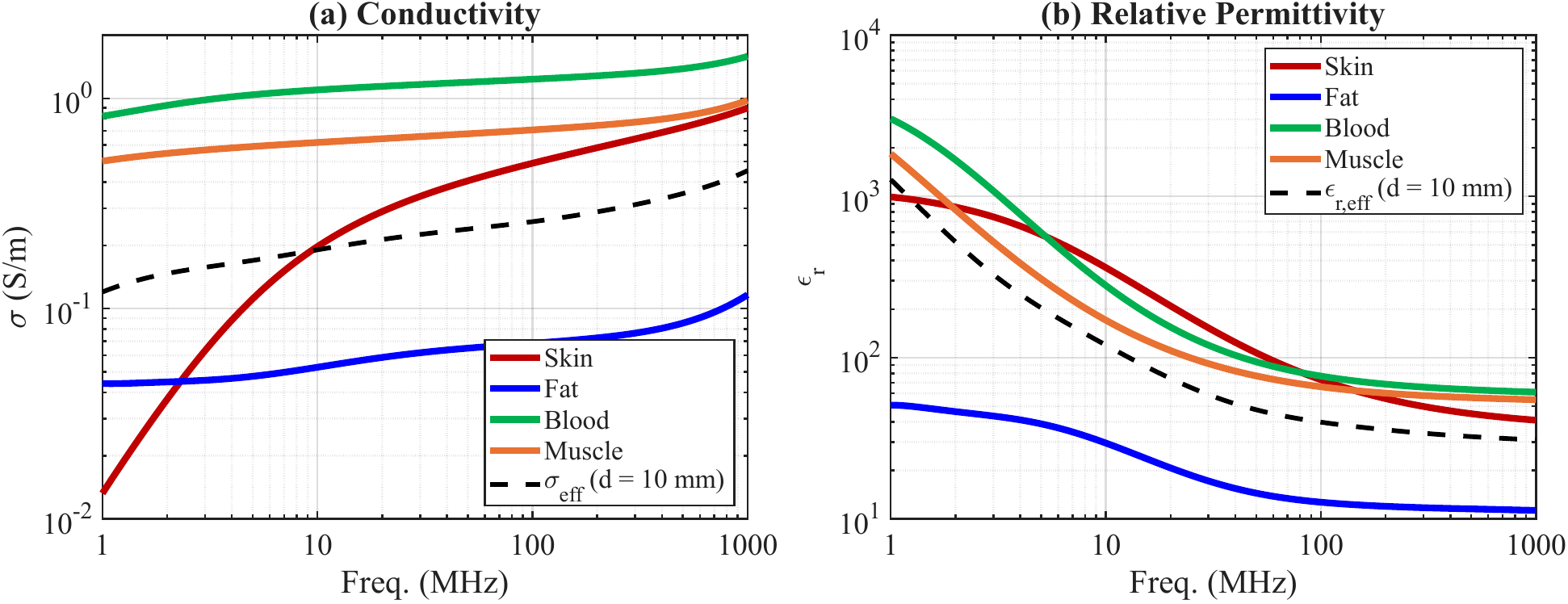}}
\caption{\red{Frequency-dependent electrical properties governing the DTCP
tissue path. (a) Conductivity and (b) relative permittivity of skin, fat,
blood, muscle \cite{gabriel1996dielectric}, and the impedance-equivalent response of the nominal
$d=10$~mm heterogeneous path.}}
\label{fig:tissue_properties}
\end{figure}

\red{For the heterogeneous tissue path, these effective properties are
frequency- and depth-dependent, with
$\varepsilon_{\mathrm{eff}}(d,f)
=\varepsilon_0\varepsilon_{r,\mathrm{eff}}(d,f)$.
In the full-wave model, the dispersive properties of skin, fat, blood,
and muscle are assigned individually using the Gabriel--Gabriel model
\cite{gabriel1996dielectric}. For interpretation of the aggregate tissue
response, the multilayer path is additionally reduced to an
impedance-equivalent cylindrical medium. Following the radial conduction
formulation for cylindrical shells \cite{Sadiku2014}, the series complex
impedances of the traversed tissues are equated to an equivalent
homogeneous path, from which $\sigma_{\mathrm{eff}}$ and
$\varepsilon_{r,\mathrm{eff}}$ are obtained. The nominal $d=10$~mm
path comprises 2-mm skin, 2-mm fat, 1-mm blood, and 5-mm muscle; the
individual layers remain explicit in the full-wave simulations.}

\red{As shown in Fig.~\ref{fig:tissue_properties}, conductivity generally
increases and relative permittivity decreases with frequency, reflecting
the dispersive tissue response. The corresponding
$\sigma_{\mathrm{eff}}$ and $\varepsilon_{r,\mathrm{eff}}$ responses
capture the aggregate electrical behavior of the nominal heterogeneous
tissue path.}

The resulting differential RX voltage is

\begin{equation}
V_{\mathrm{RX}}
=
\phi_{+}-\phi_{-}
=
-\int_{\mathbf r_-}^{\mathbf r_+}
\mathbf{E}\cdot d\mathbf{l},
\label{eq:dtcp_vrx}
\end{equation}

where $\phi_{+}$ and $\phi_{-}$ denote the electric potentials at the two
RX electrodes, $\mathbf r_{+}$ and $\mathbf r_{-}$ specify their
locations, and $d\mathbf l$ is the differential path element between
them. Hence, DTCP coupling is governed by the electric field integrated
over the RX geometry and therefore by the spatial potential difference
established across the implanted electrode pair.

\subsubsection{PTE Enhancement with \red{TX/RX Geometry Co-design}}

\red{The corresponding PTE is obtained from the differential TX and RX
ports of the full-wave model, directly relating the spatial field
distribution to the power available at the implanted receiver.}

\begin{figure}[t!]
\centerline{\includegraphics[width=1\columnwidth]
{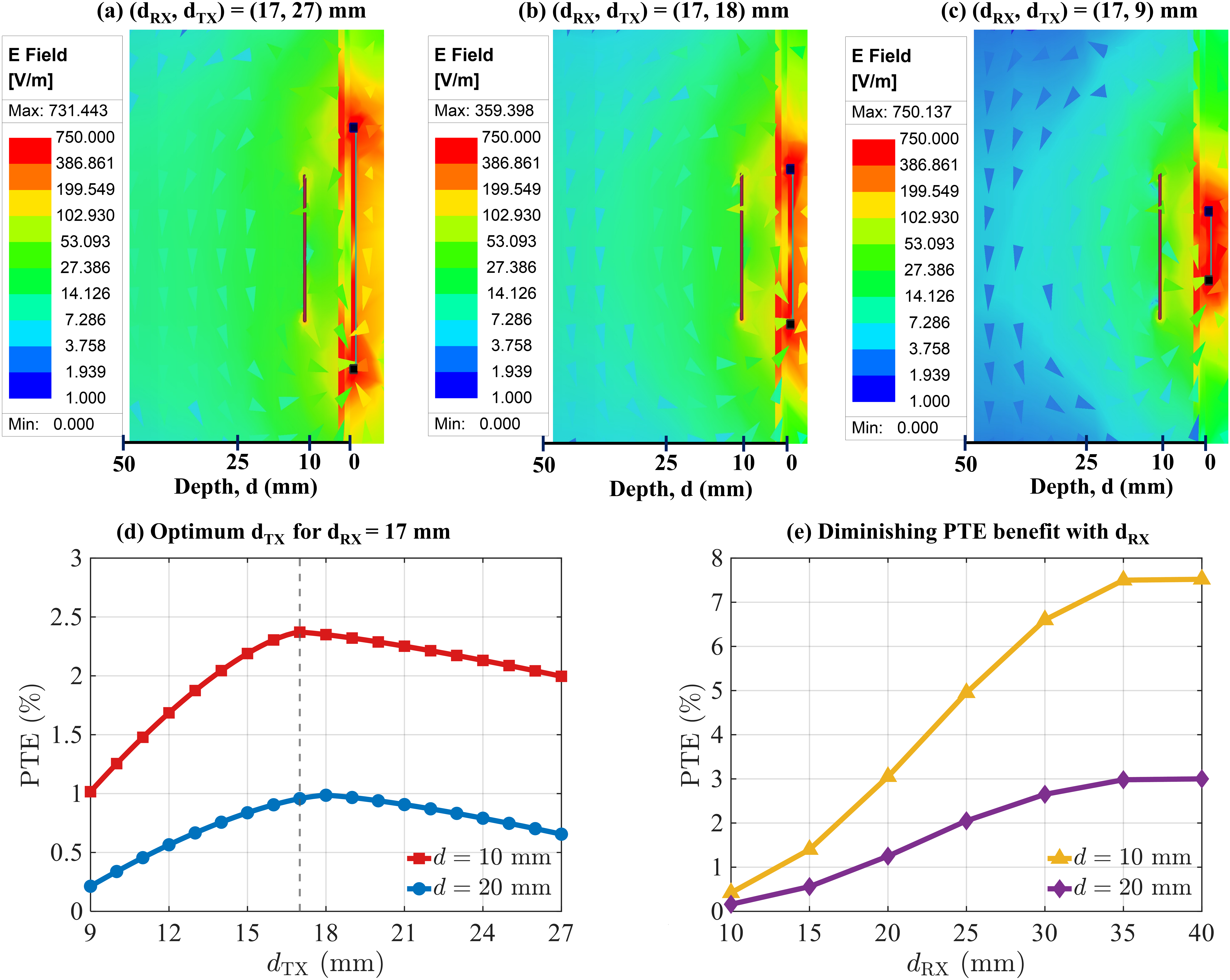}}
\caption{\red{TX/RX geometry dependence of DTCP. (a)--(c) Simulated
electric-field distributions for $d_{\mathrm{RX}}=17$~mm with
$d_{\mathrm{TX}}=27$, 18, and 9~mm. (d) PTE versus $d_{\mathrm{TX}}$
for $d=10$ and 20~mm, showing an optimum near
$d_{\mathrm{TX}}=17$--18~mm. (e) PTE versus $d_{\mathrm{RX}}$ for
$d=10$ and 20~mm, with $d_{\mathrm{TX}}$ re-optimized for each
$d_{\mathrm{RX}}$.}}
\label{fig:dtcp_field}
\end{figure}

The corresponding TX geometry must be co-optimized because
$d_{\mathrm{TX}}$ controls the launched fringing-field distribution and,
consequently, the longitudinal field gradient intercepted by the RX.
Fig.~\ref{fig:dtcp_field}\red{(a)--(c)} evaluates this dependence for
$d_{\mathrm{RX}}=17$~mm by varying $d_{\mathrm{TX}}$ from 9 to
\red{27}~mm. The resulting field redistribution establishes
$d_{\mathrm{TX}}$ as a geometry-dependent optimization variable for
maximizing differential pickup at the prescribed RX length and
implantation depth.

\red{The corresponding PTE behavior is quantified in
Fig.~\ref{fig:dtcp_field}(d). As $d_{\mathrm{TX}}$ is varied while
$d_{\mathrm{RX}}=17$~mm is fixed, PTE initially increases, reaches a
maximum near $d_{\mathrm{TX}}=17$--18~mm, and subsequently decreases
for both evaluated implantation depths. Thus, increasing
$d_{\mathrm{TX}}$ does not monotonically improve power transfer; rather,
the TX spacing must be optimized for the field distribution sampled by
the implanted RX. The observed optimum should not be interpreted as a
universal $d_{\mathrm{TX}}\approx d_{\mathrm{RX}}$ condition, since the
preferred TX spacing depends on receiver geometry and implantation
depth.}

For a thread-like receiver, the longitudinal dimension therefore becomes
an effective coupling variable. \red{As quantified in
Fig.~\ref{fig:dtcp_field}(e),} increasing the RX-electrode separation
$d_{\mathrm{RX}}$ extends the field-integration path and increases the
differential voltage available to the receiver.
This provides a direct route to higher PTE while preserving the
thread-like receiver cross section.

\red{For each $d_{\mathrm{RX}}$ in Fig.~\ref{fig:dtcp_field}(e),
$d_{\mathrm{TX}}$ is re-optimized so that the comparison reflects the
preferred TX/RX geometry rather than a fixed TX spacing. PTE increases
with $d_{\mathrm{RX}}$, while the incremental improvement diminishes at
larger receiver lengths, consistent with the nonuniform longitudinal
field sampled by the extended receiver. The lower PTE at greater
implantation depth further reflects the reduced differential excitation
available to the implanted RX.}

Thus, $d_{\mathrm{RX}}$ and $d_{\mathrm{TX}}$ provide complementary
geometric degrees of freedom: $d_{\mathrm{RX}}$ determines the
longitudinal pickup extent of the implant, whereas $d_{\mathrm{TX}}$
shapes the tissue field presented to that receiver.
\red{Together with implantation depth, these parameters therefore form
coupled DTCP design variables. In practice, the usable
$d_{\mathrm{RX}}$ is also constrained by the available anatomical
trajectory, implant flexibility, tissue conformity, packaging, and
electrode-spacing stability, and is therefore application-specific.}
\vspace{5mm}

\subsubsection{SAR Constrained Wireless Powering}

Ensuring tissue safety is essential in biomedical WPT because the maximum
permissible TX excitation, and consequently the power deliverable to the
implanted RX, is constrained by the specific absorption rate (SAR). SAR
quantifies the rate of RF-energy absorption per unit mass of tissue and
provides the basic exposure metric used to limit excessive RF-induced
tissue heating \cite{1626482,1266063,ICNIRP2020}. For frequencies between
approximately 100~kHz and 6~GHz, localized SAR constitutes the relevant
basic exposure restriction.

\red{Common public-exposure references include the FCC localized-SAR limit
of 1.6~W/kg averaged over 1~g of tissue and the ICNIRP limit of
2~W/kg averaged over 10~g for the head and torso \cite{ICNIRP2020}. Because these criteria employ different spatial
averaging masses, the corresponding SAR values should not be directly
interchanged. The analysis below uses 10-g averaged SAR and is therefore
compared with the 2-W/kg 10-g reference.}

\begin{figure}[t!]
\centerline{\includegraphics[width=1\columnwidth]
{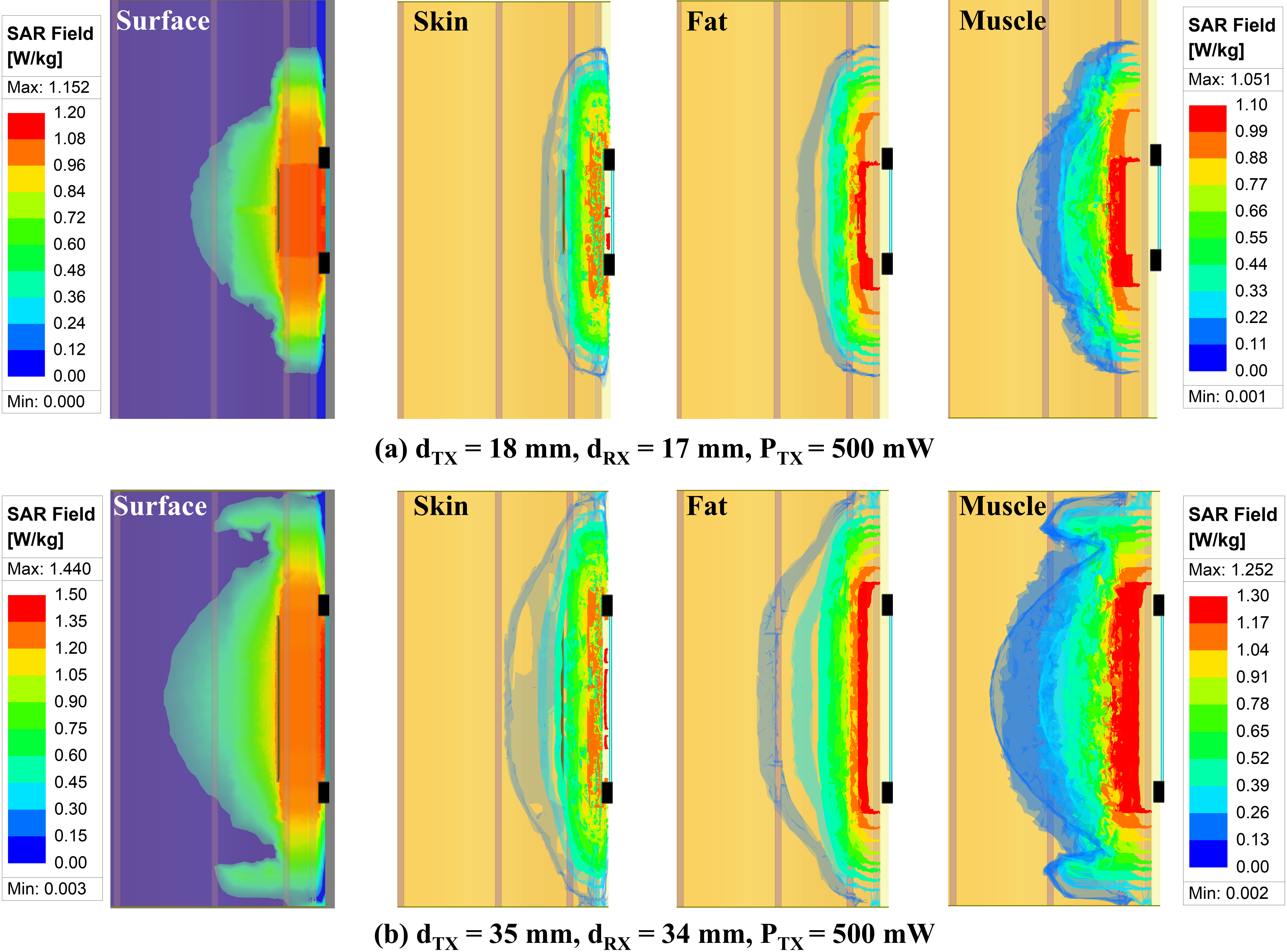}}
\caption{Simulated \red{10-g averaged} SAR distributions for $d_{\mathrm{RX}}$= 17 and 34~mm at
$P_{\mathrm{TX}}=\red{500~mW}$ \red{and $d=10$~mm}.
\red{The corresponding $d_{\mathrm{TX}}$ values are 18 and 35~mm,
respectively.} The corresponding peak SAR values are
\red{1.152 and 1.440~W/kg}, respectively, with absorption localized
around the DTCP excitation region. Both cases remain below the adopted 2-W/kg 10-g exposure reference
under the evaluated conditions.}
\label{fig:dtcp_sar}
\end{figure}

Accordingly, DTCP geometry and TX drive must be optimized under an SAR
constraint rather than from PTE alone. Fig.~\ref{fig:dtcp_sar} shows the
simulated SAR distributions at $P_{\mathrm{TX}}=\red{500~mW}$ for
$d_{\mathrm{RX}}$=\red{17 and 34}~mm. The corresponding peak SAR values
are \red{1.152 and 1.440~W/kg}, respectively, with the highest absorption
localized around the DTCP excitation region.
\red{Both configurations remain below the adopted 2-W/kg 10-g exposure
reference under the evaluated conditions.} 

\red{For a fixed geometry and tissue
model, SAR scales approximately linearly with transmitted power; therefore,
the corresponding continuous-excitation TX-power levels at which the
modeled SAR reaches 2~W/kg are approximately 0.87~W and 0.69~W for
$d_{\mathrm{RX}}=17$ and 34~mm, respectively. These values represent
continuous-excitation limits for the evaluated conditions rather than
absolute limits on instantaneous TX power. Under pulsed excitation,
instantaneous $P_{\mathrm{TX}}$ can exceed 1~W with an appropriately
reduced duty cycle, provided that the applicable time-averaged
tissue-exposure constraint remains satisfied. The SAR constraint therefore
reinforces the importance of improving geometry-dependent differential
coupling rather than relying solely on increased TX excitation.}

\red{SAR specifies absorbed RF power but does not directly quantify the
temperature rise during sustained excitation. Therefore, the maximum
simulated 10-g averaged SAR is used as the RF-heating source in a
first-order reduction of the incremental Pennes bioheat model
\cite{Pennes1948,Carluccio2013}, accounting for tissue thermal-energy
storage and perfusion-mediated heat removal:}

\begin{equation}
\red{
C_A\frac{d\Delta T}{dt}
+
B_A\Delta T
=
M_A\,\mathrm{SAR}_{10g}
}
\label{eq:thermal_ode}
\end{equation}

\red{Here, $M_A$, $C_A$, and $B_A$ denote the tissue mass, thermal
capacitance, and perfusion-mediated heat-removal coefficient per unit area,
respectively. For the heterogeneous tissue path, these are obtained from the
constituent-layer properties as
$M_A=\sum_i\rho_i h_i$,
$C_A=\sum_i\rho_i c_{p,i}h_i$, and
$B_A=\sum_i\rho_b c_{p,b}\omega_i h_i$.
The required density, specific-heat, and perfusion properties are taken
from the IT'IS tissue-property database \cite{ITIS2025}; the reported
perfusion values are converted to volumetric perfusion rates. The explicit
blood layer contributes to tissue mass and thermal capacitance but is not
assigned a separate Pennes perfusion sink.}

\red{For the nominal $d=10$~mm path comprising 2-mm skin, 2-mm fat,
1-mm blood, and 5-mm muscle, the resulting coefficients are
$M_A=10.54$~kg/m$^2$,
$C_A=3.424\times10^{4}$~J/(m$^2\!\cdot$K), and
$B_A=31.45$~W/(m$^2\!\cdot$K), corresponding to a thermal time constant
of approximately 18.15~min. The resulting transient and steady-state
temperature rises are summarized in Table~\ref{tab:thermal_response}.}

\begin{table}[t!]
\centering
\caption{\red{Predicted tissue-temperature rise under continuous
excitation at $P_{\mathrm{TX}}$=0.5 W.}}
\label{tab:thermal_response}
\resizebox{\columnwidth}{!}{
\begin{tabular}{c c c c c c}
\hline
\red{$d_{\mathrm{RX}}$} &
\red{$\mathrm{SAR}_{10g,\max}$} &
\red{$\Delta T_{10}$} &
\red{$\Delta T_{30}$} &
\red{$\Delta T_{60}$} &
\red{$\Delta T_{\mathrm{SS}}$} \\
\red{(mm)} &
\red{(W/kg)} &
\red{($^{\circ}$C)} &
\red{($^{\circ}$C)} &
\red{($^{\circ}$C)} &
\red{($^{\circ}$C)} \\
\hline
\red{17} & \red{1.152} & \red{0.164} & \red{0.312} &
\red{0.372} & \red{0.386} \\
\red{34} & \red{1.440} & \red{0.204} & \red{0.390} &
\red{0.465} & \red{0.483} \\
\hline
\end{tabular}}

\vspace{1mm}
\parbox{\columnwidth}{\footnotesize
\red{For constant SAR,
$\Delta T(t)=\Delta T_{\mathrm{SS}}(1-e^{-t/\tau_T})$, where
$\Delta T_{\mathrm{SS}}=M_A\mathrm{SAR}_{10g}/B_A$ and
$\tau_T=C_A/B_A$.}}
\end{table}
\vspace{5mm}

\red{As shown in Table~\ref{tab:thermal_response}, continuous
500-mW excitation approaches steady-state temperature rises of
0.386$^{\circ}$C and 0.483$^{\circ}$C for
$d_{\mathrm{RX}}=17$ and 34~mm, respectively. Thus, within this
first-order perfusion-aware model, the thermal response remains bounded
below 0.5$^{\circ}$C rather than increasing indefinitely. This analysis is
intended as thermal screening and does not explicitly resolve
three-dimensional heat spreading or localized vascular heat transport,
which can be incorporated using spatial bioheat models
\cite{Carluccio2013}.}

\red{Together, the link and safety analyses establish
$d_{\mathrm{RX}}$, $d_{\mathrm{TX}}$, and implantation depth as coupled
DTCP design variables, with SAR and thermal exposure setting the usable
TX-power ceiling. Exploiting the receiver's longitudinal dimension
improves differential pickup without increasing implant cross section,
while the resulting variation in tissue coupling and geometry produces a
broad received-power range at the implant. This directly motivates the
HDR energy harvester described next.}

\subsection{HDR Energy Harvester}
\begin{figure}[t!]
\centerline{\includegraphics[width=0.95\columnwidth]{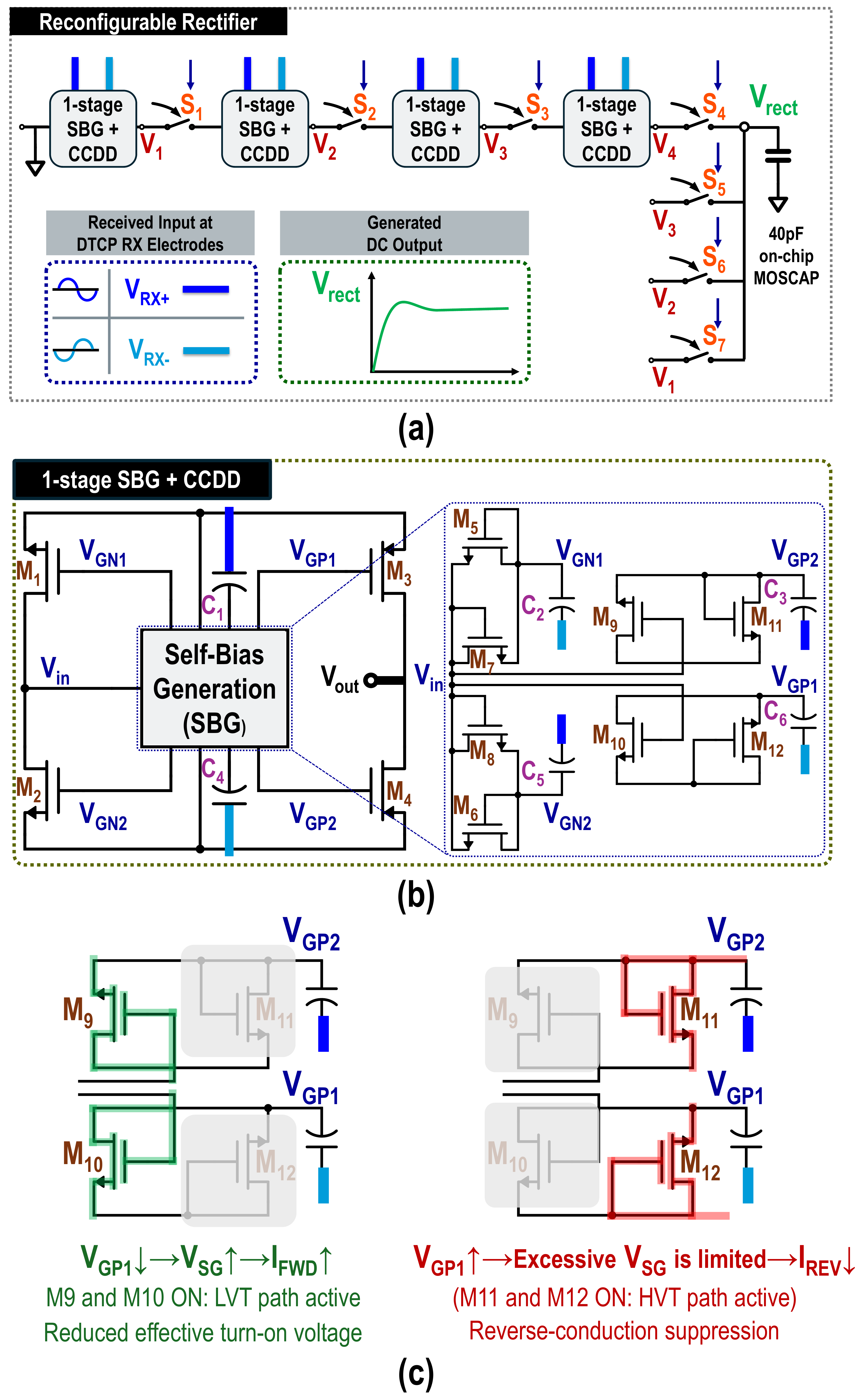}}
\caption{\red{Reconfigurable four-stage SBG-CCDD rectifier and self-bias operation. (a) Cascaded stage-selection architecture producing $V_{\mathrm{rect}}$ from $V_1$--$V_4$. (b) Single-stage SBG-CCDD with internal gate-bias generation. (c) Input-adaptive biasing, where the LVT path enhances forward conduction at low input power, while the HVT path limits excessive $V_{SG}$ and suppresses reverse conduction at high input power.}}
\label{R_SBG}
\end{figure} 

The HDR energy harvester employs the four-stage reconfigurable SBG-CCDD rectifier shown in \red{Fig.~\ref{R_SBG}(a)}. Four cascaded single-stage rectifier cells generate the intermediate rectified voltages $V_1$--$V_4$, while the switch network $S_1$--$S_7$ selects the corresponding rectifier configuration and routes the selected output to $V_{\mathrm{rect}}$. 
\red{A 40-pF on-chip MOS capacitor provides local
buffering at $V_{\mathrm{rect}}$, which also serves
as the LDO input.}
The rectifier receives the differential DTCP signals
$V_{\mathrm{RX+}}$ and $V_{\mathrm{RX-}}$.

\subsubsection{Input-Adaptive Self-Bias Generation}
\red{Fig.~\ref{R_SBG}(b)} details the single-stage SBG-CCDD cell that forms the fundamental rectifying unit of the four-stage harvester and \red{Fig.~\ref{R_SBG}(c)} illustrates the input-adaptive self-bias operation of a single SBG-CCDD stage. The differential signals $V_{RX+}$ and $V_{RX-}$ drive the cross-coupled rectifier, while the SBG locally generates $V_{GP1}$, $V_{GP2}$, $V_{GN1}$, and $V_{GN2}$. Unlike a fixed gate connection, these internally generated bias voltages vary with the received signal amplitude and therefore modify the instantaneous transistor drive without requiring an auxiliary supply, comparator, or explicit mode-control signal. This self-sustained bias adaptation follows the same fundamental wide-dynamic-range requirement identified in prior adaptive rectifiers: strong forward conduction is desirable at low input power, whereas excessive transistor drive must be restrained at higher input power to limit reverse conduction.

Considering the representative pMOS controlled by $V_{GP1}$,
\begin{equation}
V_{SG}=V_S-V_{GP1}
\end{equation}
and the corresponding overdrive is
\begin{equation}
V_{\mathrm{OV}}=V_{SG}-|V_{TP}|
\end{equation}

Under a first-order strong-inversion approximation, the drain current increases approximately as
\begin{equation}
I_D \propto \left(V_{SG}-|V_{TP}|\right)^2,
\qquad
V_{SG}>|V_{TP}|.
\end{equation}

This relation is used only to illustrate the dependence of conduction strength on the self-generated gate bias, since the transistor traverses different operating regions during a rectification cycle.

At low input power, the \red{low-threshold-voltage (LVT)} devices $M_9$ and $M_{10}$ dominate the SBG response. Their lower turn-on voltage allows the bias-generation path to become active for relatively small differential amplitudes. As illustrated in \red{Fig.~\ref{R_SBG}(c)}, this pulls $V_{GP1}$ toward a lower potential, thereby increasing $V_{SG}$ during the useful conduction interval. The larger overdrive increases $I_{\mathrm{FWD}}$ and effectively reduces the input amplitude required to establish useful rectification. This is particularly important at low received power, where limited transistor drive and forward-conduction loss dominate the conversion efficiency.

As the differential input amplitude increases, the \red{high-threshold-voltage (HVT)} devices $M_{11}$ and $M_{12}$ progressively become active, resulting in the high-power bias condition shown in \red{Fig.~\ref{R_SBG}(c)}. Their conduction shifts $V_{GP1}$ upward and prevents the pMOS $V_{SG}$ from increasing without constraint. This becomes increasingly important as the rectified DC voltage rises. During the portion of the cycle in which the instantaneous input no longer supports forward energy transfer, a strongly conducting rectifying transistor can provide a path for previously stored output energy to flow back toward the input, producing $I_{\mathrm{REV}}$. The same transistor-overdrive dependence therefore applies to the reverse-conduction interval: reducing the available $V_{SG}$ reduces channel conduction and suppresses $I_{\mathrm{REV}}$ relative to an excessively driven rectifier. Thus, the SBG does not attempt to maximize transistor overdrive over the complete input-power range; instead, it adapts the gate operating point to favor $I_{\mathrm{FWD}}$ at low input power while limiting excessive reverse conduction at higher input power.

\begin{figure}[t!]
\centerline{\includegraphics[width=1\columnwidth]{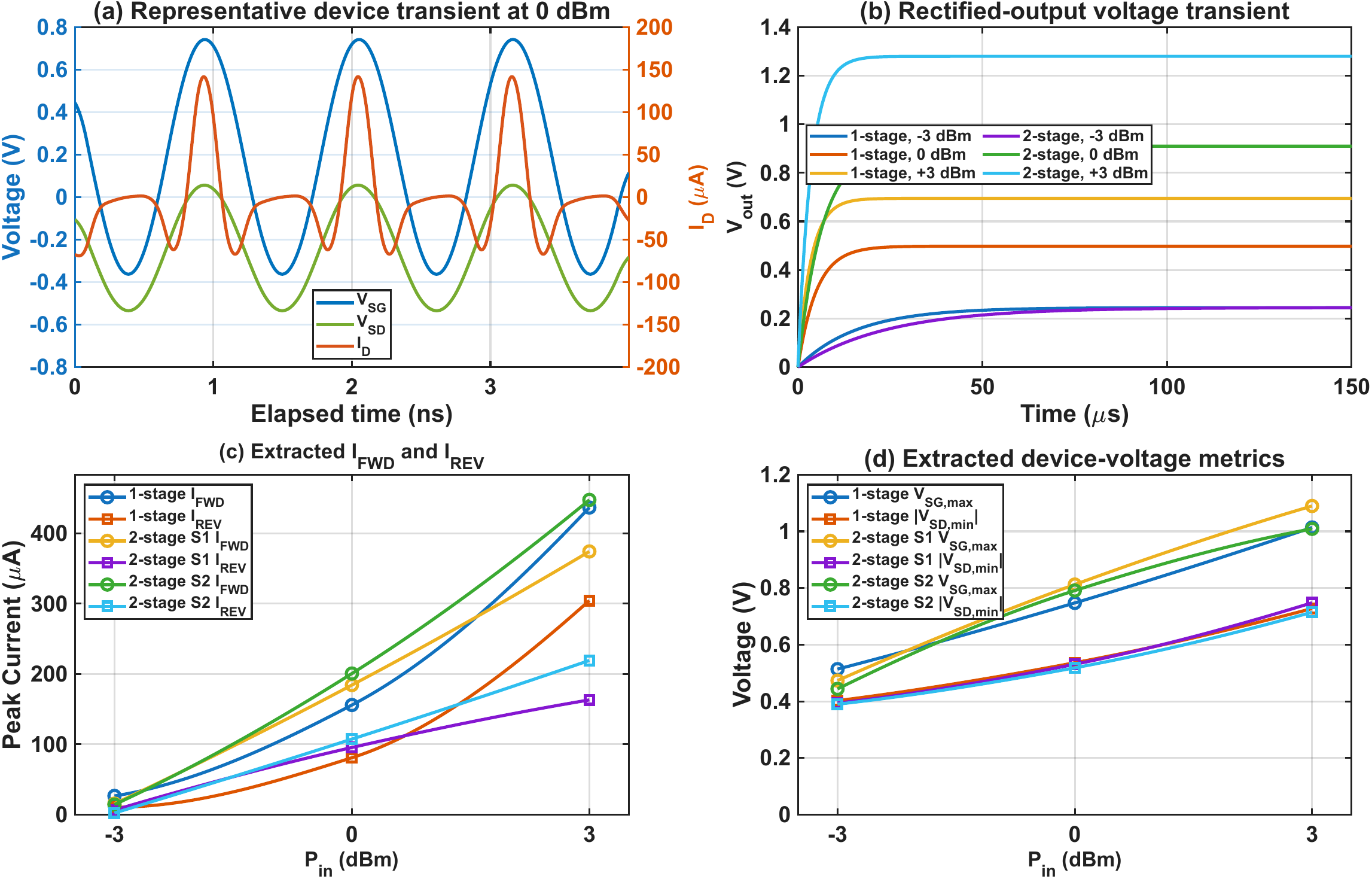}}
\caption{Characterization of the SBG-CCDD rectifier. (a) Representative $V_{SG}$, $V_{SD}$, and $I_D$ transient waveforms at 0 dBm input power. (b) Rectified-output voltage transients for the one- and two-stage configurations. (c) Extracted peak forward and reverse currents versus input power. (d) Extracted $V_{SG,\max}$ and $|V_{SD,\min}|$ for the one-stage rectifier and both stages of the two-stage rectifier.}
\label{ckt11}
\end{figure} 

The device-level behavior is further examined in Fig.~\ref{ckt11}. Fig.~\ref{ckt11}(a) shows representative $V_{SG}$, $V_{SD}$, and $I_D$ waveforms at a 0-dBm source setting for the one-stage rectifier. The current polarity is presented such that forward current is positive. A large $I_D$ pulse occurs during the forward-conduction interval when sufficient $V_{SG}$ is established, whereas the current reverses during the opposite portion of the cycle as the source-drain voltage condition changes. The significantly different forward- and reverse-current magnitudes directly illustrate the asymmetric conduction established by the SBG-CCDD.
The $-3$ to $+3$ dBm values used in Fig.~\ref{ckt11} represent the input-power
levels at the rectifier front-end applied through the balun to generate the differential signals at the rectifier input. They should therefore not be interpreted directly as the power accepted by the rectifier. Neglecting other losses, the accepted power is
\begin{equation}
P_{\mathrm{in,eff}}
=
P_{\mathrm{inc}}
\left(1-|S_{11}|^2\right),
\end{equation}

where $P_{\mathrm{inc}}$ is the incident differential power at the rectifier interface. When referenced back to the signal generator, the balun insertion loss must additionally be included. The selected source-power range is used here to expose the evolution of the internal device operation rather than to represent the complete measured input-power range of the energy harvester.

Fig.~\ref{ckt11}(b) compares the rectified-output transients of the one- and two-stage configurations. Cascading an additional SBG-CCDD stage increases the accumulated DC voltage, while each stage retains the same local self-biasing mechanism. This is verified more directly in Fig.~\ref{ckt11}(c), which extracts the peak $I_{\mathrm{FWD}}$ and $I_{\mathrm{REV}}$ for the one-stage rectifier and for both stages of the two-stage configuration. As the input amplitude increases, the available transistor drive increases and consequently $I_{\mathrm{FWD}}$ rises substantially. $I_{\mathrm{REV}}$ also becomes increasingly significant at higher input levels, highlighting the fundamental tradeoff between increased forward conduction and reverse-conduction loss. Importantly, the role of the HVT bias path is not to force the absolute $I_{\mathrm{REV}}$ to decrease with increasing input power, but rather to limit the excessive device overdrive that would otherwise further increase reverse conduction. Fig.~\ref{ckt11}(d) shows the corresponding extracted voltage quantities. $V_{SG,\max}$ increases as additional input drive becomes available, while $|V_{SD,\min}|$ also increases as the rectified DC level rises. The first and second stages of the two-stage configuration exhibit closely matched trends, demonstrating that cascading primarily translates the DC operating level while preserving the local SBG-CCDD device behavior. The two-stage analysis is therefore sufficient to demonstrate how the operating mechanism propagates through a cascade without repeating essentially identical transistor-level plots for every stage.
\begin{figure}[t!]
\centerline{\includegraphics[width=0.95\columnwidth]{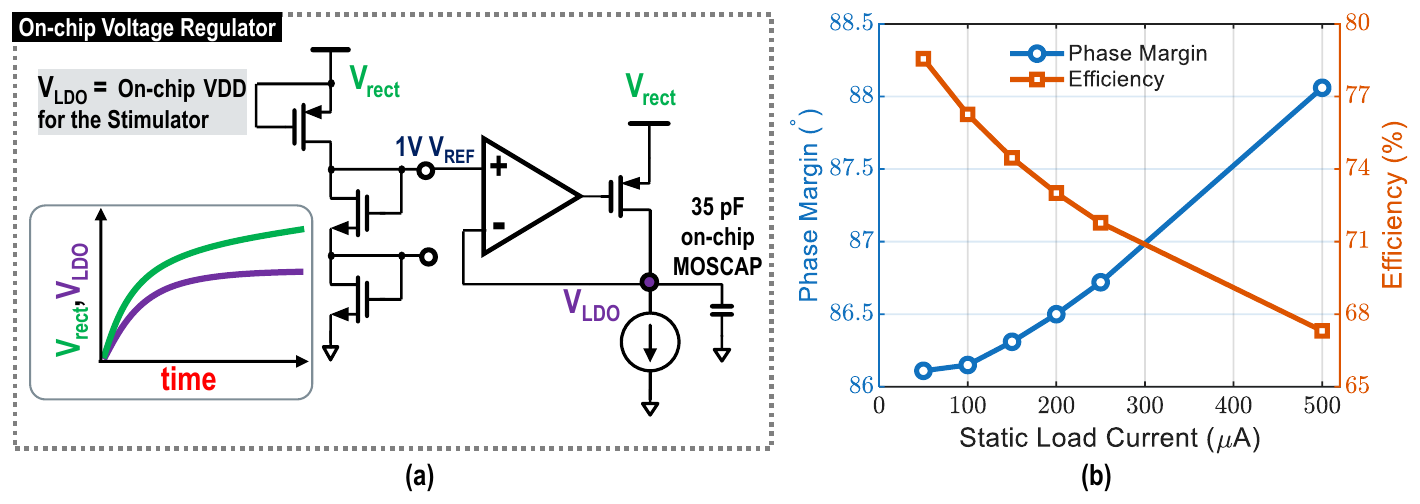}}
\caption{On-chip voltage regulator and simulated load-dependent performance.
(a) Voltage-regulation circuit that converts $V_{\mathrm{rect}}$ into the
regulated $V_{\mathrm{LDO}}$ used as the on-chip supply for the stimulator.
\red{(b) Simulated phase margin and DC efficiency versus static load current at
$V_{\mathrm{rect}}=1.2$~V.}}
\label{ckt4}
\end{figure} 

Increasing the number of active stages further increases the obtainable rectified voltage; however, each added stage also contributes additional forward-conduction loss and reverse-conduction paths. Consequently, beyond a certain stage count, the incremental DC-voltage benefit becomes progressively offset by cumulative $I_{\mathrm{REV}}$ and conduction loss, which ultimately limits PCE. The implemented four-stage architecture therefore balances voltage buildup against accumulated rectifier loss, while the reconfigurable stage-selection network allows the active stage count to be adapted across the received-power range. Each SBG-CCDD cell is therefore designed to sustain a useful high-PCE operating region, and cascading four such cells provides the intermediate DC nodes required for extending that efficient operating range through stage reconfiguration.
Also, the selected $V_{\mathrm{rect}}$ is subsequently processed by the
on-chip voltage regulator in Fig.~\ref{ckt4}. The regulator uses a 1-V reference
and a 35-pF on-chip MOS capacitor to generate $V_{\mathrm{LDO}}$, which
serves as the on-chip supply for the stimulation circuitry.
\red{The regulator employs a conventional five-transistor OTA as the error
amplifier without an additional dedicated compensation network, and
post-layout-extracted loop-gain simulations confirm stable closed-loop
operation.}
\red{At $V_{\mathrm{rect}}=1.2$~V, the phase margin remains
86.11$^\circ$--88.06$^\circ$ while the simulated DC efficiency decreases
from 78.56\% to 67.31\% as the static load current increases from
50~$\mu$A to 500~$\mu$A, as shown in Fig.~9(b).}

\subsubsection{Automatic Stage Configuring Operation}
The reconfigurable rectifier supports both manual characterization and automatic adaptive stage-selection modes. In manual mode, the comparator path is bypassed and the digital controller directly applies a valid switch code to select one of the intermediate rectified outputs $V_1$--$V_4$. For the four-stage implementation, the switch codes $0000001$, $1000010$, $1100100$, and $1111000$ select $V_1$, $V_2$, $V_3$, and $V_4$, respectively, where the bit positions correspond to $S_1$--$S_7$. This mode enables independent characterization of each rectifier-stage configuration.

\begin{figure}[t!]
\centerline{\includegraphics[width=1\columnwidth]{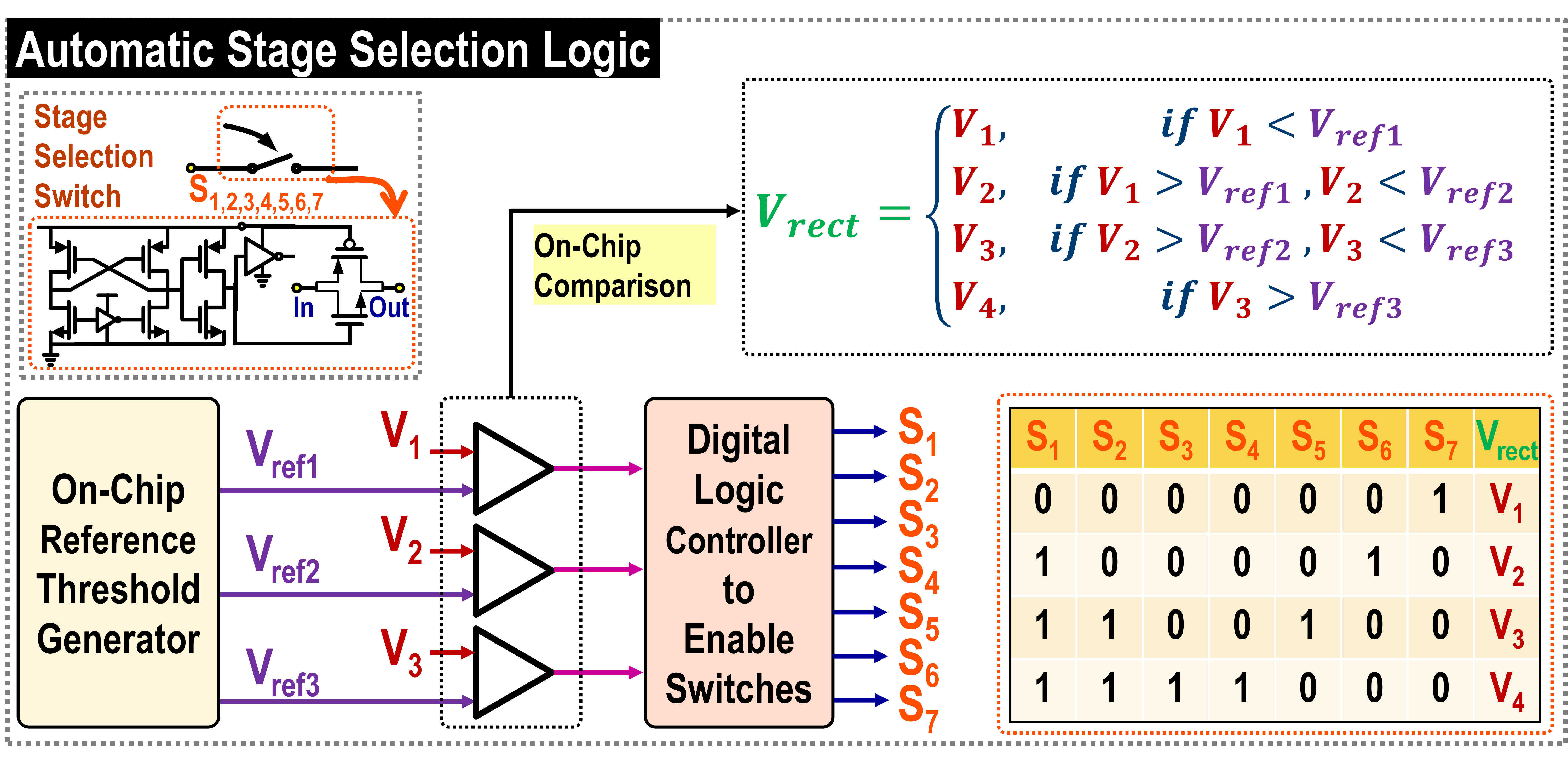}}
\caption{Automatic stage-selection logic for the reconfigurable rectifier. Intermediate stage voltages are compared with \red{PCE-derived} on-chip reference thresholds, and the digital controller generates the corresponding $S_1$--$S_7$ switch-control pattern to select the desired rectifier output stage.}
\label{ckt2}
\end{figure} 

\begin{figure}[t!]
\centerline{\includegraphics[width=1\columnwidth]{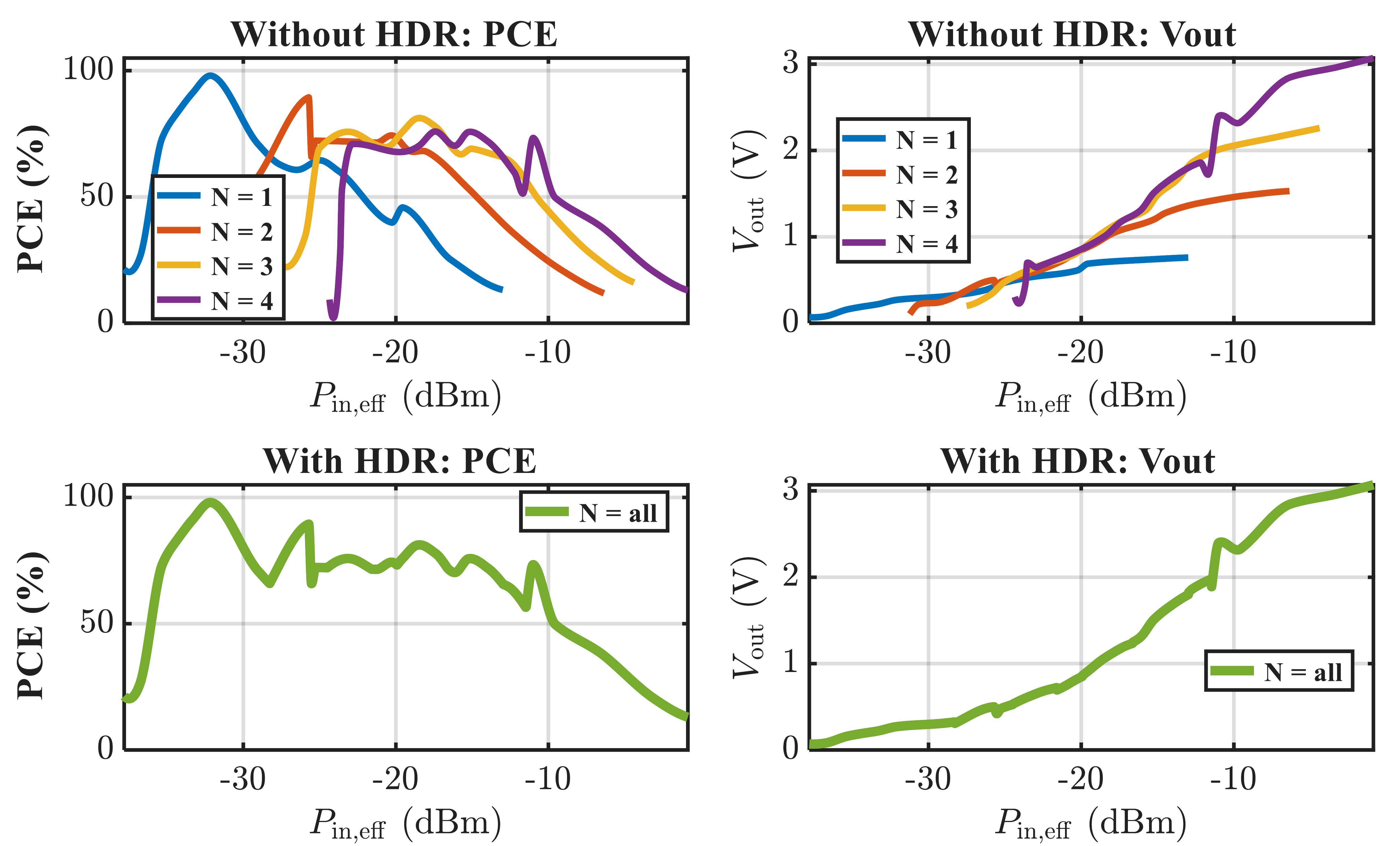}}
\caption{Simulated PCE and output-voltage characteristics versus received input power for different rectifier-stage configurations ($N=1$--$4$ and $N=\mathrm{All}$), illustrating benefits of HDR operation and how the preferred stage configuration changes with input power.}
\label{ckt5}
\end{figure} 

\begin{figure}[t!]
\centerline{\includegraphics[width=0.75\columnwidth]{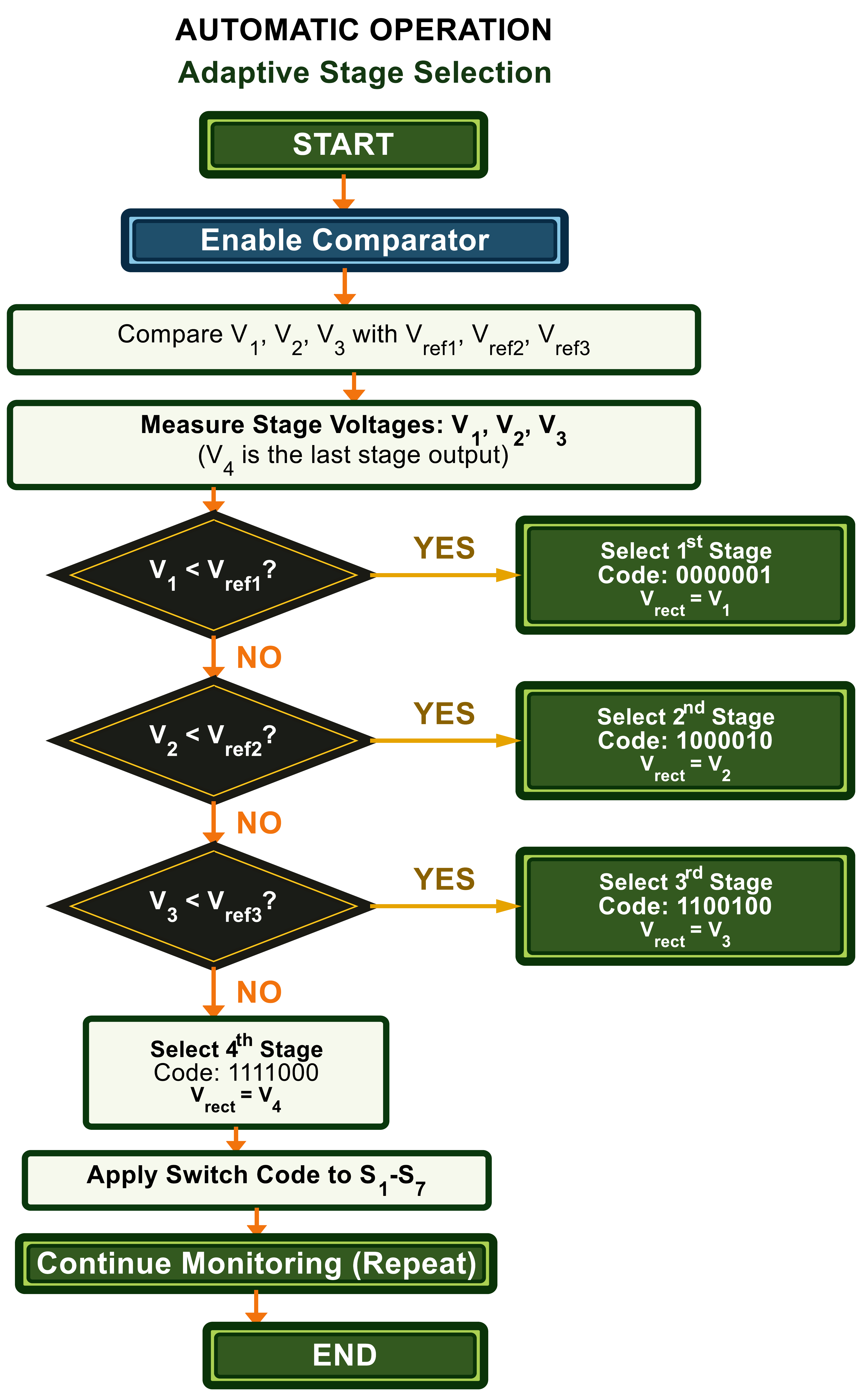}}
\caption{Flowchart of the automatic adaptive stage-selection procedure. The controller compares $V_1$, $V_2$, and $V_3$ against their respective reference thresholds, selects the corresponding rectifier stage and switch code, applies the configuration to $S_1$--$S_7$, and periodically repeats the evaluation at the update interval.}
\label{ckt6}
\end{figure} 

\red{The fixed-stage PCE characteristics are used to establish the transition references for automatic operation. Each configuration provides a finite input-power region over which comparatively high PCE is maintained, and the transition points are placed near the boundaries of these efficient operating regions. In the presented implementation, these boundaries are selected near the points where the corresponding fixed-stage PCE falls below approximately 50\%. The associated intermediate-voltage values define $V_{\mathrm{ref1}}$, $V_{\mathrm{ref2}}$, and $V_{\mathrm{ref3}}$.}
\red{The PCE used to establish these transition references is evaluated at
$V_{\mathrm{rect}}$ and does not include the low-dropout regulator (LDO) efficiency.}

\red{Across the simulated input-power range, the preferred configuration generally progresses from lower to higher stage count as the available input power increases. For a fixed-stage configuration, the increase in rectified output voltage gradually diminishes as the input power is raised. A higher-stage configuration can then provide additional rectified-voltage buildup through cascaded voltage accumulation, although at the cost of additional conduction and reverse-conduction loss. The preferred transition therefore occurs when the voltage benefit of the higher-stage configuration becomes favorable relative to its added loss, as reflected by the corresponding PCE characteristics.}

Fig.~\ref{ckt2} shows the automatic stage-selection logic. The intermediate voltages $V_1$, $V_2$, and $V_3$ are compared with the on-chip thresholds $V_{\mathrm{ref1}}$, $V_{\mathrm{ref2}}$, and $V_{\mathrm{ref3}}$, respectively, and the digital controller generates the corresponding complete $S_1$--$S_7$ switch code. \red{Here, $V_1$--$V_3$ serve as the decision variables for stage selection, while $V_1$--$V_4$ are the selectable rectified outputs.}

\red{This configuration-level adaptation complements the input-adaptive self-biasing within each SBG-CCDD stage: the self-bias network adjusts local device conduction, while the stage-selection loop determines the appropriate cascaded rectifier output for the prevailing operating condition.}
\red{The specific distinction of the proposed stage-selection scheme is therefore the use of the rectified intermediate voltages themselves as decision variables, together with transition references established from the fixed-stage PCE characteristics, to select among the available rectified outputs.}


Fig.~\ref{ckt5} shows the PCE and output-voltage characteristics of the fixed-stage configurations $N=1$--$4$ together with the automatic mode, $N=\mathrm{All}$. Each fixed configuration exhibits a distinct PCE maximum over a limited input-power region, whereas automatic operation switches among the available configurations using the predefined reference thresholds, thereby extending the overall high-efficiency operating range.

The automatic stage-selection rule is expressed as
\[
V_{\mathrm{rect}}=
\begin{cases}
V_1, & V_1<V_{\mathrm{ref1}},\\
V_2, & V_1>V_{\mathrm{ref1}},\; V_2<V_{\mathrm{ref2}},\\
V_3, & V_2>V_{\mathrm{ref2}},\; V_3<V_{\mathrm{ref3}},\\
V_4, & V_3>V_{\mathrm{ref3}}.
\end{cases}
\]

Fig.~\ref{ckt6} summarizes the automatic stage-selection procedure.
\red{A 16-bit counter initiates the evaluation once every
$2^{16}=65,536$ cycles of the 10-kHz on-chip clock, corresponding to
an update interval of approximately 6.55~s (0.153~Hz). During each
evaluation, the comparator path is enabled for two clock cycles
(200~$\mu$s), after which the comparator states associated with
$V_1$, $V_2$, and $V_3$ are used to update the corresponding
$S_1$--$S_7$ switch code. Outside this evaluation interval, the
comparator path is disabled and the selected rectifier configuration
is maintained.}

\subsection{Digitally Controlled Reconfigurable Stimulator}
\begin{figure}[t!]
\centerline{\includegraphics[width=0.9\columnwidth]{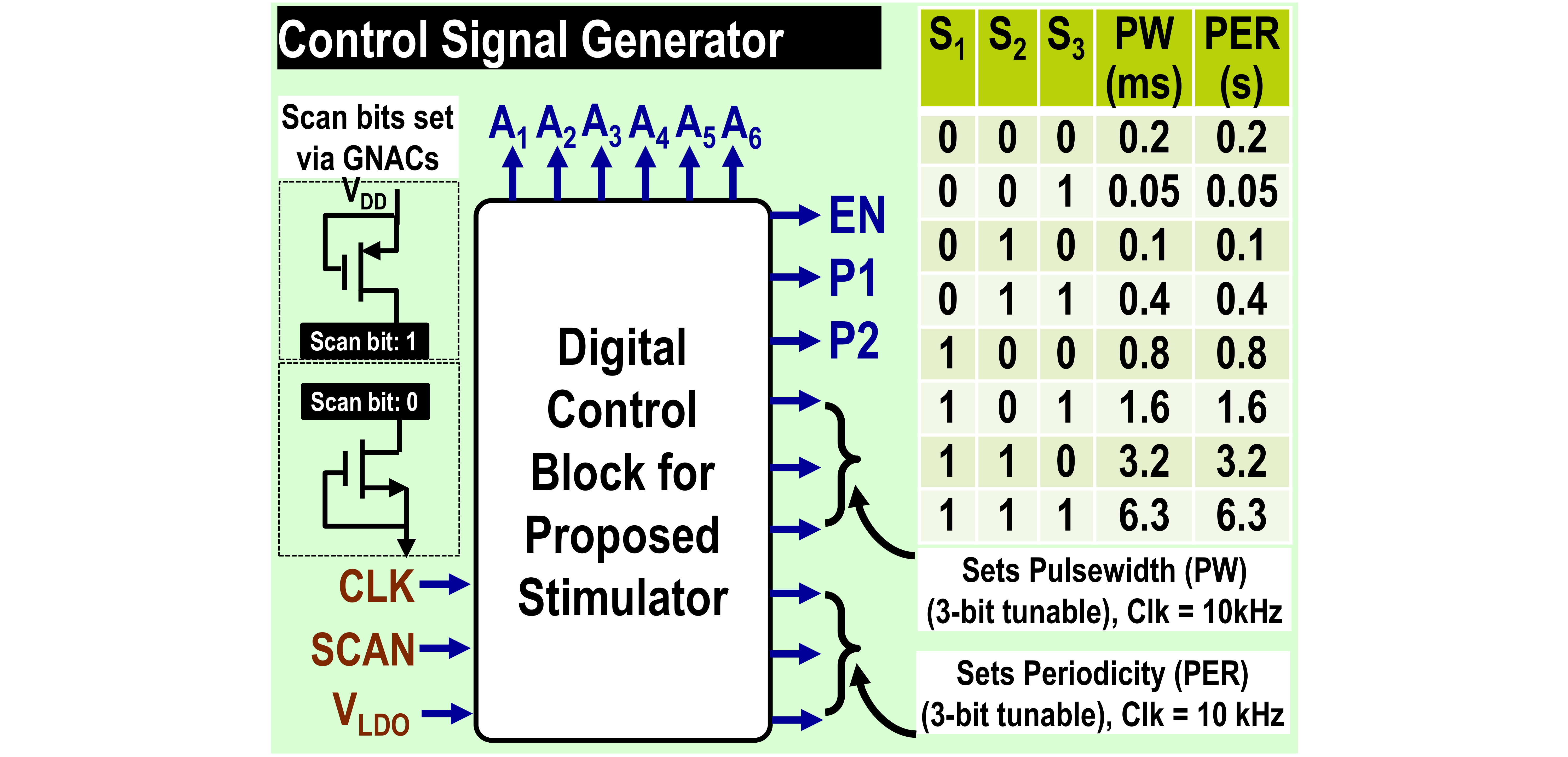}}
\caption{Digitally controlled stimulation timing generator in which scan bits configure the stimulator control block to set enable and biphasic switching signals, together with 3-bit programmable pulse width and stimulation periodicity derived from the on-chip clock.}
\label{ckt7}
\end{figure} 
\begin{figure}[t!]
\centerline{\includegraphics[width=1\columnwidth]{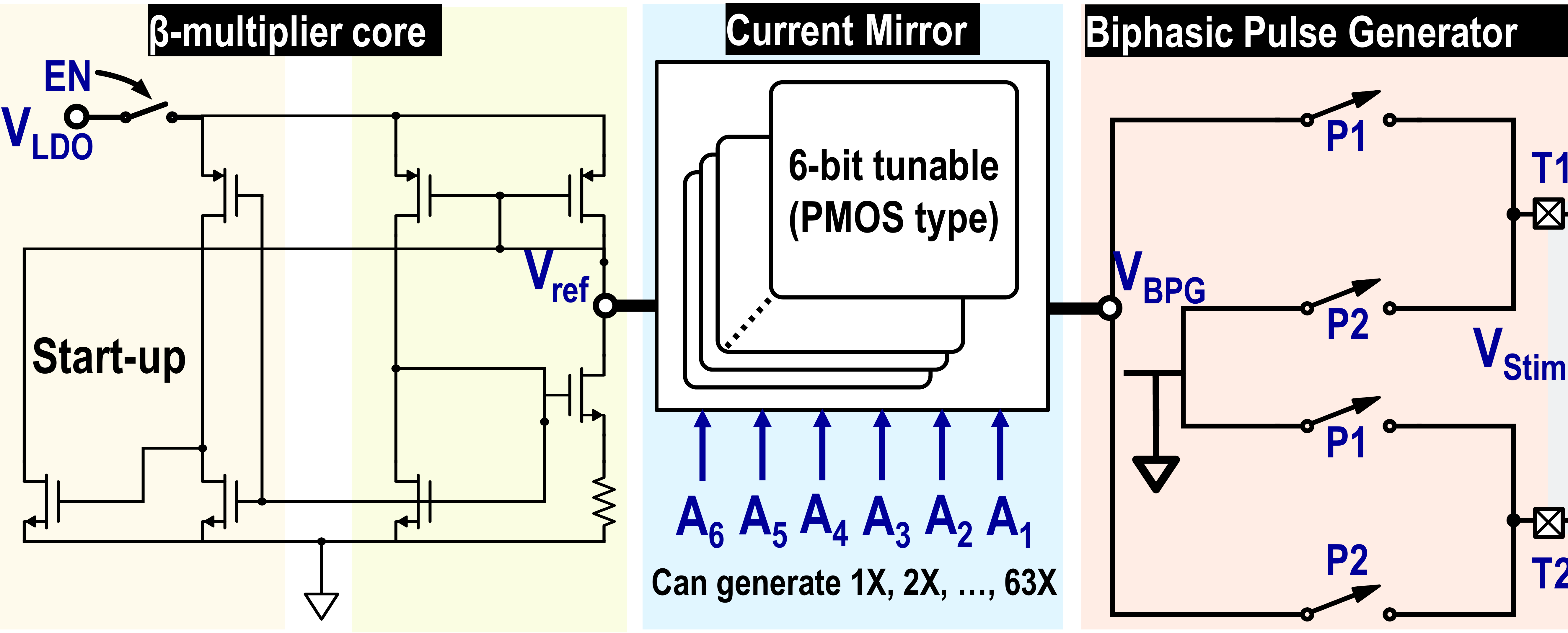}}
\caption{Tunable constant-current biphasic stimulator architecture comprising a $\beta$-multiplier reference core, 6-bit PMOS current mirror providing $1\times$--$63\times$ current scaling, and a biphasic pulse generator that steers the programmed current through the $R_L \parallel C_L$ stimulation load.}
\label{ckt8}
\end{figure} 
\begin{figure}[t!]
\centerline{\includegraphics[width=0.93\columnwidth]{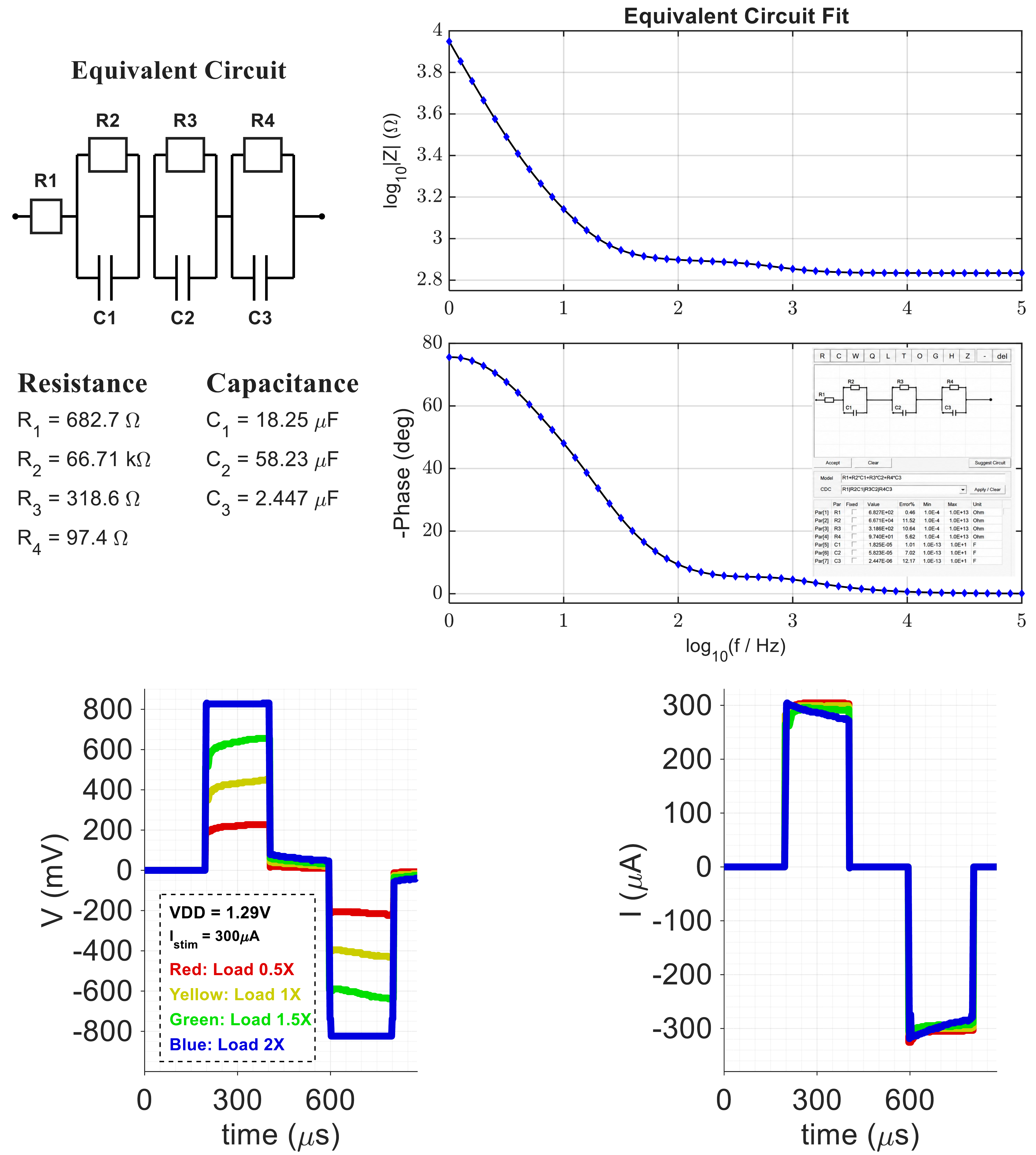}}
\caption{EIS-based stimulation-electrode load modeling and constant-current stimulator co-simulation. The electrode impedance measured from 1~Hz to 100~kHz was fitted with a multi-branch RC model and used in Cadence co-simulation. The stimulator maintains an approximately constant $300~\mu$A biphasic current across 0.5$\times$--2$\times$ load variation at $V_{DD}=1.29$~V.}
\label{ckt9}
\end{figure} 

The regulated $V_{\mathrm{LDO}}$ powers the stimulation circuitry in Figs.~\ref{ckt7} and~\ref{ckt8}.
\red{The regulated supply provides a more controlled operating voltage for
the stimulation circuitry as $V_{\mathrm{rect}}$ varies with the rectifier
operating condition, at the cost of the additional voltage drop and power
loss introduced by the LDO.}
In Figs.~\ref{ckt7}, scan-programmed control bits configure the stimulation timing from the on-chip 10-kHz clock, providing independent 3-bit control of pulse width (PW) and stimulation periodicity (PER). The programmable PW and PER scale with the applied timing clock; the ranges shown in Figs.~\ref{ckt7} correspond
to operation with the internal 10-kHz clock, whereas an external clock can
be applied during characterization to realize different timing ranges.

The constant-current stimulator in Fig.~\ref{ckt8} comprises a $\beta$-multiplier reference with start-up circuitry, a 6-bit tunable PMOS current mirror, and a biphasic pulse generator. Digital bits $A_1$--$A_6$ provide $1\times$--$63\times$ current scaling, thereby setting the stimulation amplitude, while complementary $P_1/P_2$ switching reverses the programmed current through the $R_L\parallel C_L$ load to generate the two stimulation phases. Thus, stimulation amplitude is controlled by the current-mirror code, while PW and PER are independently set by the control bits.
\red{The simultaneously sustainable combinations
of stimulation-current amplitude, pulse width,
and repetition interval depend on the available
harvested power, supply-voltage headroom, and
electrode loading.}

The stimulation-electrode impedance was characterized using an Ivium workstation through electrochemical impedance spectroscopy
(EIS) over the frequency range from 1~Hz to 100~kHz.
\red{The measured frequency-dependent impedance was fitted using the
equivalent-circuit model shown in Fig.~\ref{ckt9}, comprising a series resistance
followed by multiple parallel RC branches to represent the
electrode--electrolyte interface~\cite{Aftab2024MultiRC,Raj2022Nanoelectrodes}.
The multi-branch RC representation captures multiple relaxation time
constants, allowing the frequency dependence of both impedance magnitude
and phase to be represented more accurately than with a single RC branch.}
The fitted
impedance magnitude and phase closely follow the measured EIS response,
indicating that the extracted circuit provides an appropriate electrical
representation of the stimulation load. The extracted EIS model was subsequently imported into Cadence and co-simulated with the proposed constant-current biphasic stimulator. In addition to the nominal fitted load, scaled load conditions of
0.5$\times$, 1$\times$, 1.5$\times$, and 2$\times$ were evaluated to examine
the effect of electrode-impedance variation. For $V_{DD}=1.29$~V and a
programmed $I_{\mathrm{stim}}=300~\mu$A, the required stimulation voltage
changes substantially with load impedance, whereas the delivered biphasic
current remains close to the programmed value across the evaluated load
range. This behavior demonstrates the load-tolerant constant-current
operation of the stimulator under variations in the electrode--electrolyte
interface.

\section{Measurement Results and Discussion}

\red{Having established the DTCP link behavior, geometry dependence,
safety considerations, and the corresponding HDR energy-harvesting and
stimulation circuitry in Section~II, this section presents experimental
validation of the proposed platform at the circuit and system levels.
Building on the previously reported ex vivo demonstration of DTCP link
feasibility in~[47], the validation here proceeds in two complementary
stages. First, bench-top measurements characterize the fabricated
HDR-DTCP ASIC, including its wide-range energy-harvesting and tunable
stimulation performance. This is followed by an in vivo demonstration
using a separate injectable DTCP ASIC to verify wireless operation and
neuromuscular stimulation in a physiologically relevant setting.}

\subsection{HDR-DTCP Energy-harvester and Stimulator: CMOS Prototype and Benchtop Measurements}
\begin{figure}[t!]
\centerline{\includegraphics[width=0.95\columnwidth]{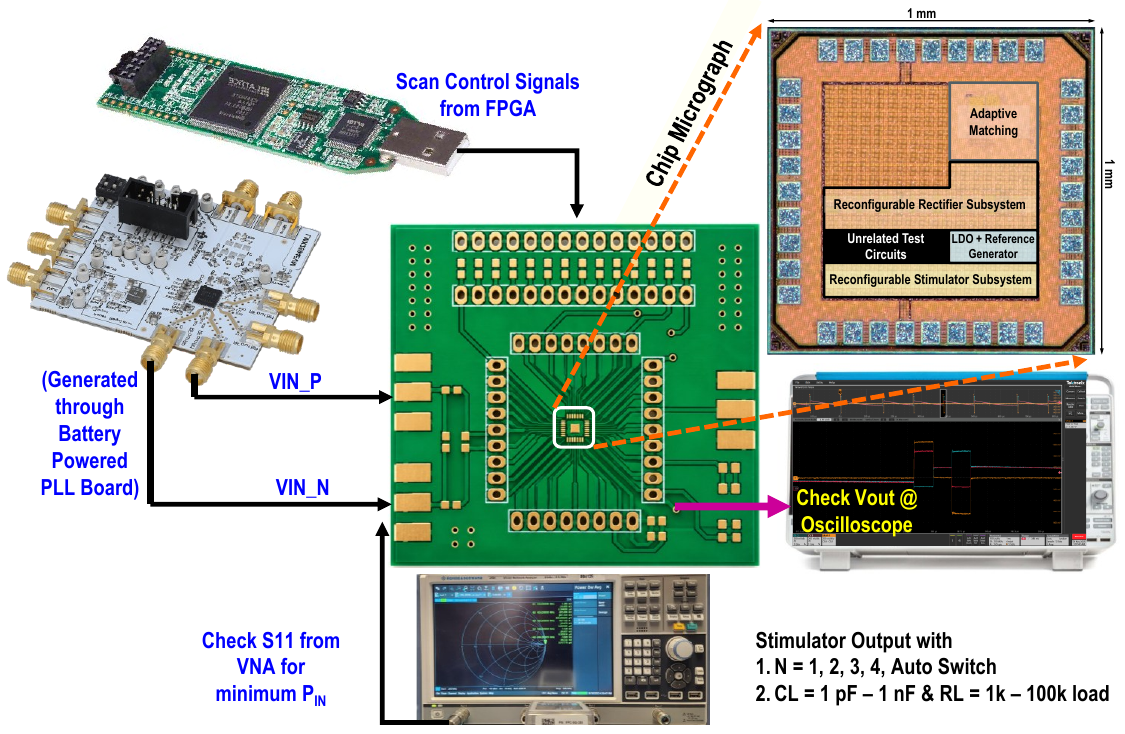}}
\caption{Measurement setup for characterization of the fabricated
$1\times1~\mathrm{mm}^2$ 65-nm CMOS prototype. The annotated die
micrograph identifies the major integrated subsystems. Differential RF
inputs are generated using a battery-powered PLL board, FPGA-generated
scan signals configure the IC, a VNA monitors the RF input condition,
and an oscilloscope records the stimulator output under programmable
rectifier configurations and electrical load conditions.}
\label{setup}
\end{figure}

\begin{figure}[t!]
\centerline{\includegraphics[width=1\columnwidth]{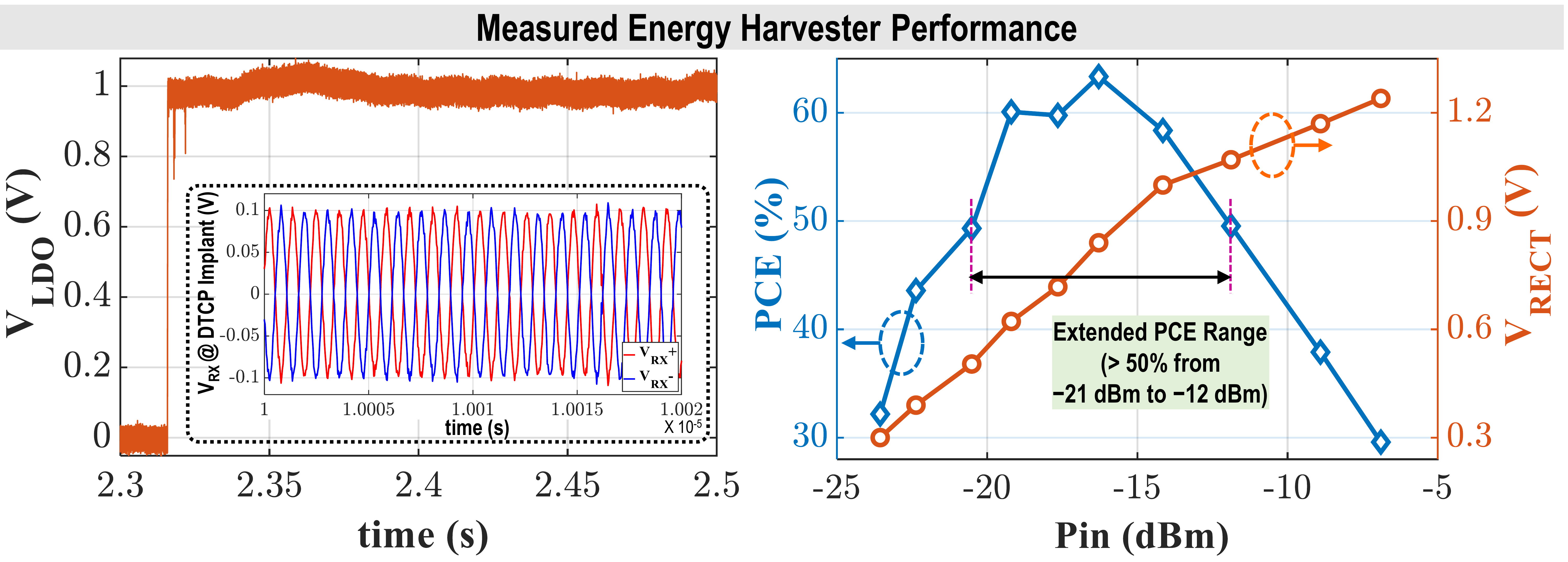}}
\caption{Measured energy-harvester performance of the fabricated
HDR-DTCP ASIC. The time-domain waveform shows the regulated output
together with the applied differential DTCP receiver voltage, while the measured
PCE and rectified voltage versus input power demonstrate $>50\%$ PCE
over the $-21$ to $-12$ dBm input-power range.}
\label{eh1}
\end{figure}

\begin{figure}[t!]
\centerline{\includegraphics[width=0.75\columnwidth]{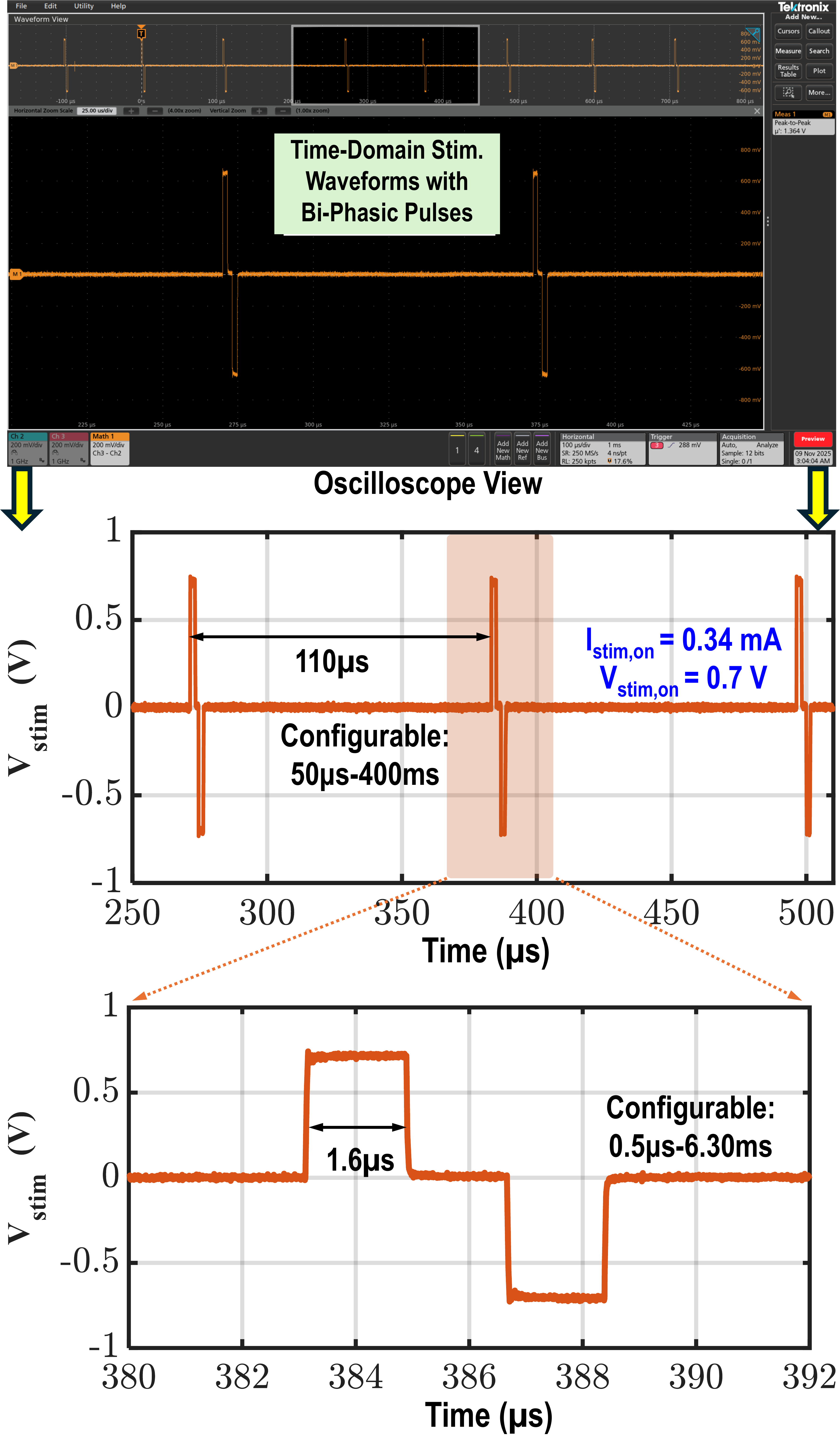}}
\caption{Measured time-domain biphasic stimulation waveforms. At $V_{\mathrm{LDO}}=0.8$~V and a $2~\mathrm{k}\Omega\parallel5~\mathrm{pF}$ load, the stimulator delivers $0.34$~mA with approximately $0.7$~V amplitude; the enlarged waveform illustrates a 1.6-$\mu$s biphasic pulse with 110-$\mu$s periodicity and programmable pulse-width and repetition-period ranges.}
\label{stim1}
\end{figure}

The fabricated 65-nm CMOS prototype occupies
$1\times1~\mathrm{mm}^{2}$ and integrates the reconfigurable rectifier,
adaptive matching network, on-chip low-dropout regulator (LDO) and
reference-generation circuitry, and the reconfigurable stimulation subsystem.
Fig.~\ref{setup} shows the annotated die micrograph together with the
benchtop setup used for electrical characterization of the fabricated IC.
A battery-powered \red{phase-locked loop (PLL)} board generates
the differential RF inputs, $V_{\mathrm{IN,P}}$ and $V_{\mathrm{IN,N}}$, while an external \red{field-programmable gate array (FPGA)} provides the scan-control signals required to program the rectifier and stimulator configurations. The RF input condition is monitored using a vector network analyzer (VNA), and the corresponding rectified, regulated, and stimulation outputs are measured using an oscilloscope. The setup allows the effective number of rectifier stages to be varied from one to four, including the automatic switching mode, while the stimulator is evaluated across different resistive and capacitive load conditions. This enables independent characterization of the energy-harvesting and stimulation paths prior to
integration with the tissue-coupled link.
\red{During bench-top measurements, an additional
2.2-$\mu$F off-chip capacitor was connected at
$V_{\mathrm{rect}}$ to reduce supply-voltage
droop during pulsed loading.}

\begin{figure*}[t!]
\centerline{\includegraphics[width=1\textwidth]{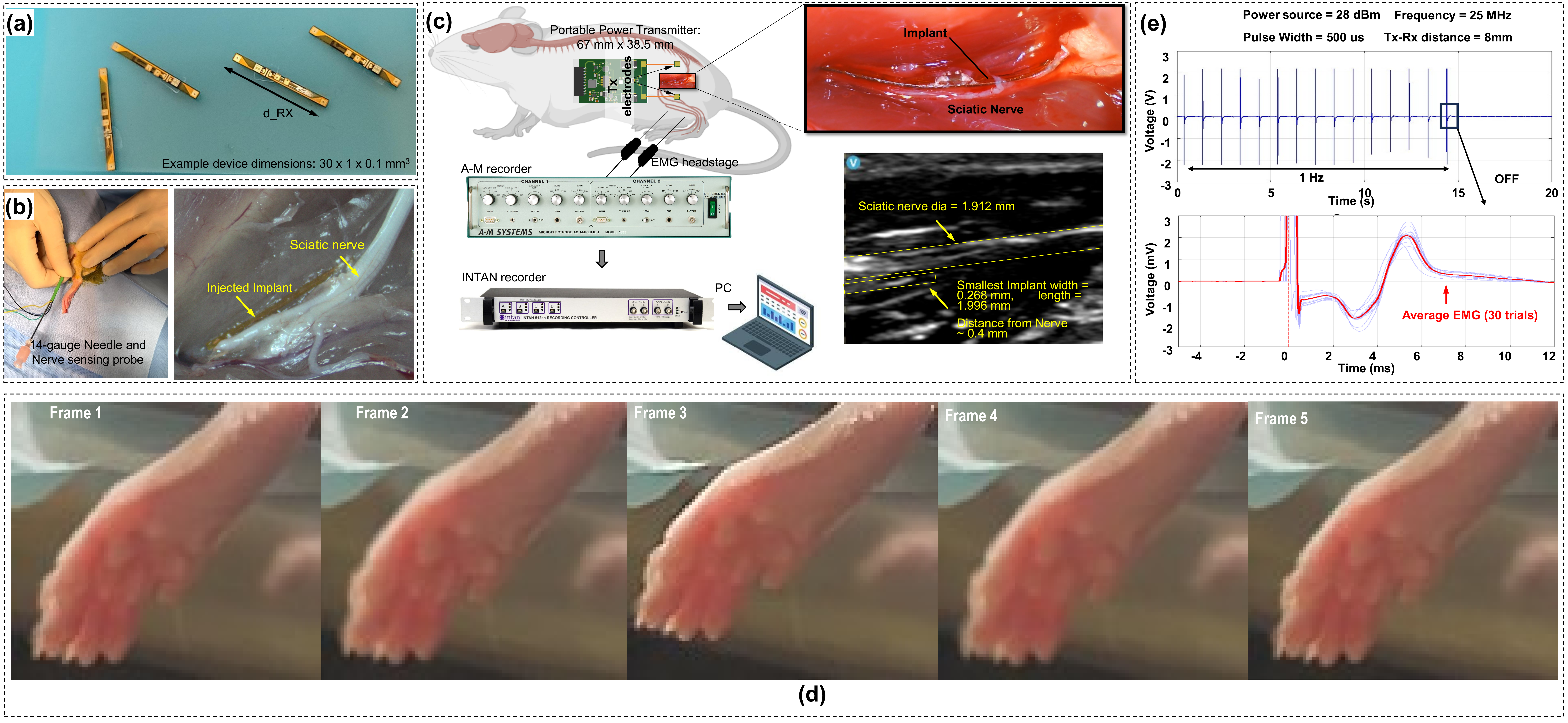}}
\caption{\textit{In vivo} DTCP with an injectable ASIC-based implant \cite{Han_adv_sci} : (a) Thread-like devices with ASIC assembled; (b) Device injection with a 14-gauge needle and a Nerve-sensing probe, as shown in \cite{Han_adv_sci, han2025ner}; (c) Measurement setup for wireless stimulation of rodent sciatic nerve. A portable DTCP transmitter drives the skin on the hind legs, using two external TX electrodes, while EMG signals are recorded from the biceps femoris muscle. The implant is placed near the sciatic nerve - Ultrasound imaging confirms placement of the implanted device; (d) DTCP-powered stimulation of the rat sciatic nerve, with resulting hind-limb twitch visualized over five sequential video frames and paw-tip motion tracking showing twitch progression under a 500-$\mu$s/s stimulation pattern; (e) Wireless stimulation elicits consistent, repeatable EMG responses. The representative EMG trace demonstrates time-locked muscle activation synchronized with each 500-$\mu$s/s stimulation pulse.}
\label{invivo}
\end{figure*}

The measured energy-harvester characteristics are presented in
Fig.~\ref{eh1}. The differential DTCP receiver signal applied to the ASIC input is
rectified and regulated to establish an approximately 1-V supply for
 subsequent stimulation circuitry.
As the received input power
is varied, the reconfigurable rectifier achieves a peak PCE of 63.5\% at
$P_{\mathrm{in}}=-16.3$~dBm and maintains PCE above 50\% from approximately
$-21$ to $-12$~dBm, corresponding to a 9-dB high-efficiency input-power
range. Over this region, the rectified voltage increases with the available
input power while remaining sufficient for subsequent voltage regulation. The broad high-PCE region demonstrates the principal advantage of the
reconfigurable energy harvester: rather than optimizing the rectifier for a
single received-power condition, the effective rectifier configuration can be
adapted to preserve efficient AC-to-DC conversion over a substantially wider
input-power range. 
The regulated time-domain waveform further confirms that the applied
DTCP receiver signal can be converted into a stable supply for subsequent
implant electronics.

Fig.~\ref{stim1} presents the measured time-domain response of
the reconfigurable biphasic stimulator. For $V_{\mathrm{LDO}}=0.8$~V and a
$2~\mathrm{k}\Omega\parallel5$~pF load, the stimulator delivers
$I_{\mathrm{stim,on}}=0.34$~mA, corresponding to an output amplitude of
approximately 0.7~V. The measured waveform exhibits sequential positive and
negative phases, providing the biphasic stimulation required for
charge-balanced excitation. The representative measurement uses a 1.6-$\mu$s pulse width and a 110-$\mu$s stimulation interval. Both timing parameters are programmable: the pulse width can be configured from approximately 0.5~$\mu$s to 6.30~ms, while the stimulation interval can be varied from 50~$\mu$s to 400~ms. \red{These measurements verify that the integrated stimulator can
independently control the stimulation timing while preserving the biphasic
output waveform under controlled electrical loading. The
$2~\mathrm{k}\Omega \parallel 5~\mathrm{pF}$ condition in Fig.~\ref{stim1}
represents one measured operating condition; additional
$R_L \parallel C_L$ conditions were used during benchtop characterization,
whereas the EIS-derived multi-branch model in Fig.~\ref{ckt9} was used separately
to evaluate the constant-current response under the measured
frequency-dependent electrode--electrolyte impedance and its variation.}

\subsection{Preliminary \textit{in-vivo} Injectable DTCP Demonstration}
Complementing the bench-top characterization
of the fabricated HDR-DTCP ASIC, Fig.~19
demonstrates needle-based injectability and
DTCP-powered neural stimulation using a separate
injectable DTCP ASIC \cite{Han_adv_sci}.
A DTCP-ASIC-based implant (energy harvester + stimulator) is assembled on a narrow
flexible substrate together with the DTCP receiving and stimulation electrodes, resulting in a thread-like device suitable for needle-based delivery (Fig.~\ref{invivo}(a)).
To test injectability, a dummy implant was delivered adjacent to the rat sciatic nerve using a nerve-sensing probe integrated into a 14-gauge spinal needle, with real-time nerve proximity confirmed via deep-learning-assisted signal processing on the probe's sensing electrodes prior to implant placement, similar to \cite{Han_adv_sci}, as shown in Fig.~\ref{invivo}(b). Saline hydrodissection created a pocket adjacent to the nerve for implant delivery via push rod, after which the needle was withdrawn and the injection tract sealed with alginate hydrogel. Ultrasound imaging confirmed implant placement approximately 0.4 mm from the sciatic nerve, and post-placement dissection corroborated accurate positioning along the nerve trajectory.

To show stimulation capabilities, in a separate experiment (thread-like DTCP devices surgically placed, and not injected, to have better control over the device placement and experiment), wireless excitation was provided by a portable, battery-powered DTCP transmitter (67 mm × 38.5 mm, 3.7 V lithium supply) \cite{Han_adv_sci}, via copper micro-pin TX electrodes driven at a MHz-band carrier frequency optimized per animal (n=4),
\red{with the TX power varied during the experiment up to a maximum available RF output of 28~dBm for the portable DTCP transmitter (Fig.~\ref{invivo}(c))}.
Following DTCP powering, the implanted stimulator generated a 500 $\mu$s stimulation waveform (configurable) delivered to the nerve at 1 Hz. The resulting hind-limb motor response was visualized using five sequential video frames, with paw-tip motion tracking capturing the temporal progression of the evoked twitch (Fig.~\ref{invivo}(d)). The \red{electromyography (EMG)} signals recorded from the biceps femoris muscle showed clear, time-locked evoked potentials (Fig.~\ref{invivo}(e)) coincident with each stimulation pulse, and increasing TX power correspondingly increased both the stimulation amplitude and EMG response magnitude, confirming RF-power-dependent neural activation. Control trials with the implant removed but the same TX waveform applied produced only stimulus-locked electrical artifacts with no corresponding EMG response, ruling out direct field stimulation. 
 This experiment shows that the thread-like ASIC-based implant can be minimally invasively positioned near the target nerve, wirelessly powered through the DTCP link, and used to elicit a reproducible, quantifiable in-vivo stimulation response. All procedures were approved by the \textit{Institutional Animal Care and Use Committees of the University of Florida and Harvard University (IACUC protocols 202500000323 and 22-10-428, respectively)}.

\subsection{Effect of Implant Length}
\begin{figure}[t!]
\centerline{\includegraphics[width=0.8\columnwidth]{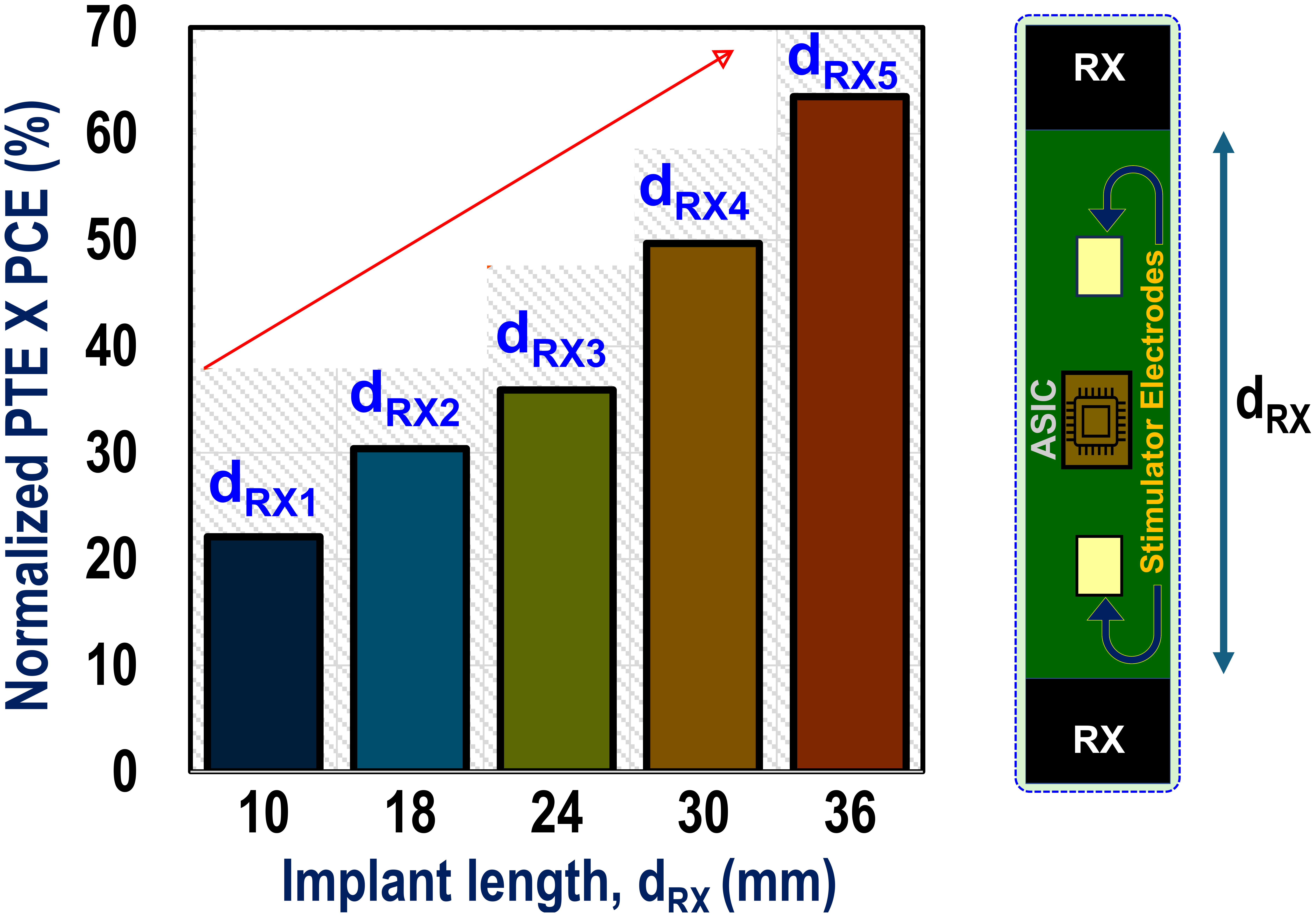}}
\caption{\red{Effect of RX electrode separation together with HDR rectifier on improving the combined $\eta_{\mathrm{coupling}}\eta_{\mathrm{rectifier}}$, resulting in higher normalized PTE$\times$PCE for longer thread-like implants.}}
\label{eta}
\end{figure}

The dependence of the end-to-end power-delivery performance on implant length
is experimentally evaluated in Fig.~\ref{eta}. As
$d_{\mathrm{RX}}$ is increased from 10 to 36~mm, the measured normalized
PTE$\times$PCE increases monotonically, with the 36-mm configuration providing
approximately three times the value obtained with the shortest implant. This
trend originates from the differential nature of DTCP, for which the voltage
developed across the RX electrodes is determined by the electric-field
integral along the implant.
Therefore, for an approximately uniform local electric field, extending
$d_{\mathrm{RX}}$ increases the differential excitation available to the
energy harvester without requiring a corresponding increase in implant
cross-sectional area. The measurement consequently verifies an important scaling property of DTCP:
power-transfer performance can be enhanced primarily by extending the
thread-like implant along its longitudinal dimension, preserving the small
cross section required for needle-based delivery.

\subsection{Comparison with State-of-the-Art}
\begin{table*}[t]
\centering

\caption{Comparison with State-of-the-Art Implantable RX for Wireless Powering and Stimulation}
\label{comparison}

\renewcommand{\arraystretch}{1.18}
\setlength{\tabcolsep}{3.0pt}

\resizebox{\textwidth}{!}{%
\begin{tabular}{l c c c c c c c c c c c c c c}

\hline

& \multicolumn{4}{c}{\textbf{WPT}}
& \multicolumn{7}{c}{\textbf{Energy Harvester}}
& \multicolumn{3}{c}{\textbf{Stimulator}}
\\

\cline{2-15}

\textbf{Work}
&
\makecell{\textbf{Link}\\\textbf{Architecture}}
&
\makecell{\textbf{Chip Size}\\\textbf{(mm$^2$)}}
&
\textbf{Validation}
&
\textbf{Injectable}
&
\textbf{Technique}
&
\makecell{\textbf{Process}\\\textbf{(nm)}}
&
\makecell{\textbf{Freq}\\\textbf{(GHz)}}
&
\makecell{\textbf{$R_{\mathrm{LOAD}}$}\\\textbf{(k$\Omega$)}}
&
\makecell{\textbf{PDR}\\\textbf{(dB)}}
&
\makecell{\textbf{PCE$_{\mathrm{max}}$ (\%)}\\
\textbf{@ $P_{\mathrm{in}}$ (dBm)}}
&
\makecell{\textbf{\# of}\\\textbf{Stages}}
&
\makecell{\textbf{Max. Stim.}\\\textbf{Power (mW)}}
&
\makecell{\textbf{Stim.}\\\textbf{Efficiency (\%)}}
&
\makecell{\textbf{Max $I_{\mathrm{stim}}$ (mA)}\\
\textbf{/ \# Ctrl bits}}
\\

\hline


\textbf{This work}
&
\textbf{HDR-DTCP}
&
\textbf{$1\times1$}
&
\makecell{\textbf{Bench-top}\\\textbf{In-Vivo$^{*}$}}
&
\textbf{Yes$^{*}$}
&
\makecell{\textbf{Reconfig.}\\\textbf{SBG + CCDD}}
&
\textbf{65}
&
\textbf{0.9}
&
\textbf{20}
&
\textbf{9}
&
\textbf{63.5, $-16.3$}
&
\textbf{4}
&
\textbf{0.9}
&
\textbf{83.2}
&
\textbf{0.9/6}
\\

\hline


\makecell{\textbf{ISSCC'20}\\\textbf{\cite{yu2020implantable}}}
&
ME
&
$1\times1.5$
&
In-Vitro
&
No
&
\makecell{CCDD +\\Comparator}
&
180
&
$<0.001$
&
N/A
&
N/A
&
N/A
&
N/A
&
1.5
&
90
&
1.5/5
\\

\hline


\makecell{\textbf{CICC'25}\\\textbf{\cite{liao2025implantable}}}
&
US
&
$1\times1.5$
&
In-Vivo
&
No
&
\makecell{Reconfig.\\Charge Pump}
&
65
&
$<0.001$
&
1.5
&
N/A
&
73.1, N/A
&
N/A
&
2.2
&
71.7
&
1.2/6
\\

\hline


\makecell{\textbf{VLSI'21}\\\textbf{\cite{chatterjee2021biphasic}}}
&
BP-QBC
&
$1\times1$
&
In-Vitro
&
No
&
\makecell{Cascaded\\Dickson}
&
65
&
0.9
&
1
&
N/A
&
$<10$, N/A
&
30
&
1.2
&
89.2
&
1/6
\\

\hline


\makecell{\textbf{CICC'25}\\\textbf{\cite{zou2025parallel}}}
&
ME
&
$1.4\times1.67$
&
Bench-top
&
No
&
\makecell{Switched\\Cap + SSHI}
&
180
&
N/A
&
N/A
&
N/A
&
60, N/A
&
1
&
-
&
-
&
-
\\

\hline


\makecell{\textbf{TVLSI'23}\\\textbf{\cite{choo2023high}}}
&
RF Antenna
&
N/A
&
Bench-top
&
No
&
Dual Topology
&
130
&
0.9
&
100
&
4
&
78.4, $-16$
&
3
&
-
&
-
&
-
\\

\hline


\makecell{\textbf{TCAS II'23}\\\textbf{\cite{choo2023high2}}}
&
RF Antenna
&
N/A
&
Bench-top
&
No
&
\makecell{Hybrid\\Topology}
&
65
&
0.9
&
100
&
6
&
79.8, $-18$
&
3
&
-
&
-
&
-
\\

\hline

\end{tabular}%
}

\vspace{3pt}
\begin{minipage}{\textwidth}
\footnotesize
\raggedright
$^{*}$Injectability and in vivo validation refer to the DTCP device demonstrated in Fig.~\ref{invivo}.
\end{minipage}

\end{table*}


Table~\ref{comparison} compares the proposed HDR-DTCP receiver with representative
state-of-the-art wireless-powering and stimulation systems across the WPT
interface, energy harvester, and stimulator. The DTCP receiver further exploits
its longitudinal dimension for power pickup without increasing cross section,
which is favorable for needle-based implantation. \red{At the system level,
the DTCP platform is the only entry in Table~\ref{comparison} with demonstrated
injectability, complementing the circuit-level energy-harvesting and stimulation
functions with an in vivo demonstration of a thread-like DTCP implant.}

At the energy-harvester level, the four-stage reconfigurable SBG-CCDD rectifier
maintains $>50\%$ PCE over a 9-dB input-power range. By comparison, the
dual-topology and hybrid-topology inductive rectifiers in~\cite{choo2023high,choo2023high2}
provide 4- and 6-dB ranges, respectively, under the $>50\%$-PCE criterion.
The proposed receiver also incorporates programmable biphasic stimulation within
the same receiver architecture. Overall, the comparison highlights the main
distinction of HDR-DTCP: length-scalable tissue coupling, wide-range rectification,
tunable stimulation, and in-vivo injectability are addressed jointly rather than
as separate subsystem demonstrations.

\section{Conclusion}
HDR-DTCP establishes a system-level route toward injectable stimulators in which receiver geometry, energy harvesting, and stimulation are co-designed rather than treated as independent blocks. The key outcome is the ability to exploit the longitudinal dimension of a thread-like implant for tissue-coupled power transfer while preserving efficient AC-to-DC conversion over a wide received-power range and maintaining programmable constant-current stimulation under load variation.
\red{Preliminary in-vivo results support the feasibility of needle-based deployment and wireless neural stimulation through DTCP.}
Although HDR-DTCP mitigates the major link- and load-dependent effects targeted in this work, biological and deployment-related uncertainties remain for chronic operation (\textit{e.g.}, encapsulation degradation, electrode stability, fibrotic response, battery-free duty-cycling under variable posture etc.). 
\red{In particular, fibrotic response and chronic evolution of the
electrode--tissue interface can alter the received operating condition of
the implanted receiver.}
Future implementations can therefore focus on link-aware harvester control, further reduction of implant volume and peripheral overhead, and longer-term in vivo evaluation. Ultimately, extending HDR-DTCP toward closed-loop and multi-node systems will require the wireless link, power-management path, and neural interface to adapt cooperatively to the evolving implant environment.

\section*{Acknowledgment}
Research reported in this publication
was supported in part by the NSF CAREER Award ECCS \#2441610, in part by the NIH NIBIB Award R01EB039082, and in part by the NIH NIBIB Award DP2EB037188.


\bibliographystyle{IEEEtran}
\bibliography{main}

\end{document}